\documentclass[preprintnumbers,10pt,nofootinbib]{revtex4}
\usepackage{amsmath,latexsym,amssymb,amsfonts}
\usepackage{color,graphicx}
\usepackage{bm}
\usepackage{subfloat}
\usepackage{caption}
\usepackage{subfig}

\begin{document}

\title{Primordial Gravitational Waves and Rescattered Electromagnetic Radiation in the Cosmic Microwave Background}

\author{Dong-Hoon Kim$^{1}$ and Sascha Trippe$^{2,3}$}

\affiliation{
$^{1}$Basic Science Research Institute, Ewha Womans University, Seoul 03760, Republic of Korea; ki13130@gmail.com\\
$^{2}$Department of Physics and Astronomy, Seoul National University, Seoul 08826, Republic of Korea; trippe@astro.snu.ac.kr\\ 
$^{3}$Corresponding author
}

\begin{abstract}
Understanding the interaction of primordial gravitational waves
(GWs) with the Cosmic Microwave Background (CMB) plasma is important for
observational cosmology. In this article, we provide an analysis of an effect apparently overlooked as yet. We consider a single
free electric charge and suppose that it can be agitated by primordial GWs
propagating through the CMB plasma, resulting in periodic, regular motion
along particular directions. Light reflected by the charge will be partially
polarized, and this will imprint a characteristic pattern on the CMB. We
study this effect by considering a simple model in which anisotropic
incident electromagnetic (EM) radiation is rescattered by a charge sitting
in spacetime perturbed by GWs and becomes polarized. As the charge is driven
to move along particular directions, we calculate its dipole moment to
determine the leading-order rescattered EM radiation. The Stokes parameters
of the rescattered radiation exhibit a net linear polarization. We
investigate how this polarization effect can be schematically represented
out of the Stokes parameters. We work out the representations of gradient
modes (E-modes) and curl modes (B-modes) to produce polarization maps.
Although the polarization effect results from GWs, we find that its
representations, the E- and B-modes, do not practically reflect the GW
properties such as strain amplitude, frequency and polarization states.
\end{abstract}

\maketitle
%\tableofcontents
%\newpage

\section{Introduction\label{intro}}

Ever since its experimental detection \cite{penzias1965} and subsequent
interpretation as a relic of the earliest epoch of the universe \cite%
{dicke1965}, studies of the Cosmic Microwave Background (CMB) have been
crucial for constraining cosmological models. Present-day space-based
measurements of the angular power spectrum of the CMB temperature
distribution on the sky are able to constrain the parameters included in the 
$\Lambda $CDM model of cosmology \cite{bahcall1999} with statistical
uncertainties down to a few per cent \cite{planck2014}. Of special
importance is the fact that state-of-the-art cosmological observations can,
at least in principle, place constraints on the large family of models of
cosmic inflation \cite{martin2014} which is a cornerstone of hot big-bang
cosmology.

A key prediction made by models of inflation is the occurrence of primordial
gravitational waves (GWs). Even though not observable directly (currently),
they should be detectable indirectly via a characteristic linear
polarization of the CMB \cite{polnarev1985}. CMB polarization can be
decomposed into contributions from gradient modes (E-modes) and curl modes
(B-modes), with the latter ones being excited by either tensor perturbations, i.e., propagation of primordial GWs through the plasma emitting the CMB, or gravitational lensing by foreground matter \cite{zaldarriaga1997,
kamionkowski1997}. Polarimetric observations \cite{trippe2014} of the CMB at
radio frequencies around 100~GHz have been numerous, with different classes
of observations aimed at different polarization modes (E or B) and angular
scales. A first observation of E-mode polarization was achieved by the
Degree Angular Scale Interferometer in 2002 \cite{kovac2002}. B-mode
polarization due to gravitational lensing was first detected by the South
Pole Telescope in 2013 \cite{hanson2013} and has since been observed by POLARBEAR \citep{polarbear2014, polarbear2015}, ACTPol \citep{act2014}, and BICEP1 \citep{bicep12014}. A detection of B-mode polarization
caused by primordial GWs was claimed by the BICEP2 collaboration in 2014 
\cite{bicep2014} but was shown to be likely caused by interstellar dust soon
thereafter \cite{flauger2014}.

The aforementioned studies aim at comparisons of theoretical polarization --
specifically B-mode -- patterns to observational ones in order to constrain
cosmological quantities. In a nutshell, the polarization patterns studied so far originate as follows: GWs propagating through the CMB plasma cause local
quadrupole anisotropies in the radiation field; this in turn causes
characteristic polarization patterns in the light scattered at CMB plasma
electrons. However, at least a priori one should also expect polarization
immediately from interactions between GWs and individual electric charges.
An electric charge located in the path of propagation of a GW will be forced
into an oscillatory motion by the wave. This should lead to
characteristic linear polarization of light scattered by the charge. We
herewith present what appears to be the first quantitative analysis of this effect. We work out the representations of gradient modes
(E-modes) and curl modes (B-modes) to produce maps of the polarization
patterns resulting from scattering at a single electric charge. Throughout this work, we adopt the negative metric signature $(-,+,+,+)$, and the Minkowski metric is given by $\eta _{\mu \nu}=\mathrm{diag}\left(-1,1,1,1\right) $.

\section{Rescattered EM radiation and polarization in the CMB \label{rescat}}

\subsection{EM radiation rescattered by a charge in perturbed spacetime 
\label{EM}}

Consider an electric charge $q$ with mass $m$ in the CMB plasma. If the
charge encounters incident light, it reradiates (outgoing) reflected light.
If, however, the charge is set in oscillating motion by a GW, the geometry
of reflection will be given by the GW: as the charge is driven to move along
particular directions, like the GW polarizations $h_{+}$ or $h_{\times }$,
it will reradiate the reflected light along the same directions.

The electromagnetic (EM) radiation rescattered by the charge, as measured by
a distant observer, can be obtained by solving the Maxwell equations 
\begin{equation}
\square A_{\mu }=-\frac{4\pi }{c}j_{\mu },  \label{ME}
\end{equation}%
where $\square \equiv -\partial ^{2}/c^{2}\partial t^{2}+\partial
^{2}/\partial x^{2}+\partial ^{2}/\partial y^{2}+\partial ^{2}/\partial
z^{2} $ denotes the d'Alembertian, with $c$ representing the speed of light, and $A_{\mu }$
represents the EM vector potential produced by the charge and $j_{\mu }$ the
charge's current density. A solution to the Maxwell's equations can be
expressed to the leading-order as 
\begin{equation}
A_{\mu }\sim \frac{\dot{Q}_{\mu }\left( t_{R}\right) }{r},  \label{A}
\end{equation}%
where $r$ is the distance from the charge to the observer and $t_{R}\equiv t-%
{r}/{c}$ denotes the retarded time, and $Q_{\mu }$ is the charge-dipole
moment defined as 
\begin{equation}
Q_{\mu }=qX_{\mu }  \label{Q}
\end{equation}%
for the charge $q$ at position $X_{\mu}$, and the overdot denotes differentiation with respect to time $t$.

Our rescattered EM radiation can be computed using Eqs. (\ref{A}) and (\ref
{Q}), with the charge's trajectory $X_{\mu }\left( t\right) $ determined in
the spacetime%
\begin{equation}
g_{\mu \nu }=\eta _{\mu \nu }+h_{\mu \nu },  \label{metric}
\end{equation}%
where $\eta _{\mu \nu }$ is the flat background and $h_{\mu \nu }$ describes
the perturbations by GWs.\footnote{%
In this work, to simplify our analysis, the spacetime is given by the metric 
$ds^{2}=$ $\left( \eta _{\mu \nu }+h_{\mu \nu }\right) dx^{\mu }dx^{\nu }$
rather than $ds^{2}=a^{2}\left( \tau \right) \left[ -d\tau ^{2}+\left(
\delta _{ij}+h_{ij}\right) dx^{i}dx^{j}\right] $, the
Friedmann-Robertson-Walker metric for the expanding universe, where $a\left(
\tau \right) $ denotes the scale factor, with $\tau $ being the conformal
time. Here we assume that during the epoch of reionization the universe was
expanding at a rate substantially slower than the characteristic frequency
of the CMB radiation, $\sim 10^{11}~\mathrm{Hz}$. Therefore, to a good
approximation, our EM radiation from a `single' charge can be well-described
in the perturbed flat spacetime.} As the charge is moving in curved
spacetime, to treat the general relativistic effects of the charge's motion
we employ a semi-relativistic approximation \cite{Ruffini}, in which the
charge's trajectory $X_{\mu }\left( t\right) $ is identified with a geodesic
in the spacetime $g_{\mu \nu }$. For a weakly perturbed spacetime, in which $%
\left\vert h_{\mu \nu }\right\vert \ll 1$, one may split the dipole moment, 
\begin{equation}
Q^{\mu }=qX_{\mathrm{(flat)}}^{\mu }+q\mathcal{X}^{\mu },  \label{Q1}
\end{equation}%
and the vector potential accordingly, 
\begin{equation}
A^{\mu }=A_{\mathrm{(flat)}}^{\mu }+\mathcal{A}^{\mu },  \label{A1}
\end{equation}%
where the perturbation $\mathcal{A}^{\mu }$ to the potential $A_{\mathrm{%
(flat)}}^{\mu }$ results from the deviation $\mathcal{X}^{\mu }$ from the
trajectory $X_{\mathrm{(flat)}}^{\mu }$ which is caused by the perturbation
of the flat background $\eta _{\mu \nu }$ by GWs $h_{\mu \nu }$.

\begin{figure}[tbp]
\centering
\includegraphics[scale=0.40]{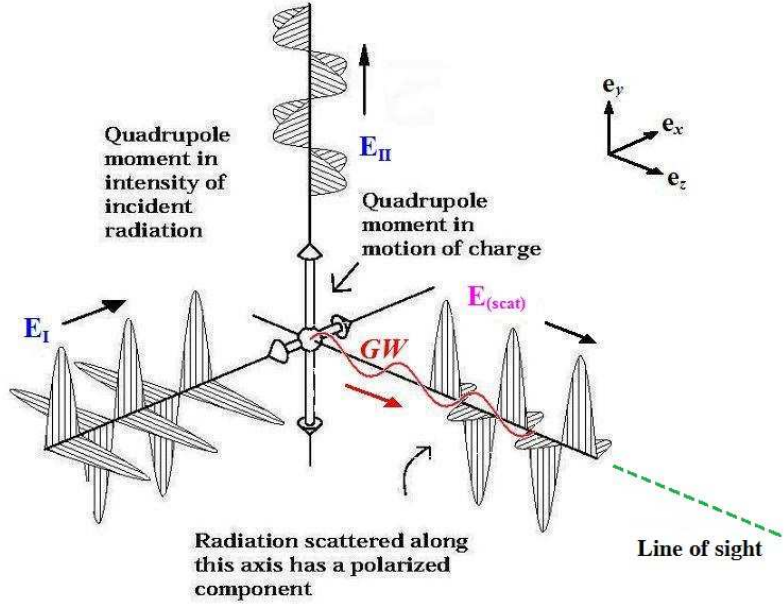}
\caption{An anisotropic incident EM radiation is rescattered by a charge
sitting on spacetime ripples (i.e. GWs) and becomes polarized. The GWs
propagate along the line of sight while being polarized in a plane
perpendicular to the the line of sight, and the rescattered (outgoing) EM
radiation is induced by the GWs. Note that the directions of propagation and
polarization of the GWs are fitted into the observer's reference frame; with
propagation along the $z$-axis (coinciding with the line of sight) and
polarization along the $x$-axis and the $y$-axis. (Credit: CAPMAP website, 
\texttt{http://quiet.uchicago.edu/capmap}; reproduced with some modifications)}
\label{fig1}
\end{figure}

Now, we prescribe an anisotropic incident monochromatic radiation field on a
charge as shown in Figure \ref{fig1} by means of the two waves,%
\begin{eqnarray}
\mathbf{E}_{\mathrm{I}\,(\mathrm{in})} &=&E_{\mathrm{I}}\left\{ \mathbf{e}%
_{y}\exp \left[ \mathrm{i}\left( Kx-\Omega t+\varphi _{\mathrm{I}}\left(
t\right) \right) +\mathbf{e}_{z}\exp \left[ \mathrm{i}\left( Kx-\Omega
t+\vartheta _{\mathrm{I}}\left( t\right) \right) \right] \right] \right\} ,
\label{Ei} \\
\mathbf{E}_{\mathrm{II}\,(\mathrm{in})} &=&E_{\mathrm{II}}\left\{ \mathbf{e}%
_{x}\exp \left[ \mathrm{i}\left( Ky-\Omega t+\varphi _{\mathrm{II}}\left(
t\right) \right) \right] +\mathbf{e}_{z}\exp \left[ \mathrm{i}\left(
Ky-\Omega t+\vartheta _{\mathrm{II}}\left( t\right) \right) \right] \right\}
,  \label{Eii}
\end{eqnarray}%
where $E_{\mathrm{I}}$ and $E_{\mathrm{II}}$ ($E_{\mathrm{I}}\neq E_{\mathrm{%
II}}$) are the amplitudes, and $\Omega $ and $K$ are the angular temporal
frequency and angular spatial frequency with $\Omega =cK$, and $\varphi _{%
\mathrm{I}}\left( t\right) $, $\vartheta _{\mathrm{I}}\left( t\right) $, $%
\varphi _{\mathrm{II}}\left( t\right) $ and $\vartheta _{\mathrm{II}}\left(
t\right) $ are the phase modulation functions, which are assumed to vary on
a time scale much slower than the period of the waves, i.e. $\left\vert \dot{%
\varphi}_{\mathrm{I}}\right\vert $, $\left\vert \dot{\vartheta}_{\mathrm{I}%
}\right\vert $, $\left\vert \dot{\varphi}_{\mathrm{II}}\right\vert $, $%
\left\vert \dot{\vartheta}_{\mathrm{II}}\right\vert \ll \Omega $. Here each
of $\mathbf{E}_{\mathrm{I}\,(\mathrm{in})}$ and $\mathbf{E}_{\mathrm{II}\,(%
\mathrm{in})}$ is \textit{unpolarized}: as the relative phases $\varphi _{%
\mathrm{I}}\left( t\right) -\vartheta _{\mathrm{I}}\left( t\right) $ and $%
\varphi _{\mathrm{II}}\left( t\right) -\vartheta _{\mathrm{II}}\left(
t\right) $ fluctuate in time, each wave will not remain in a single
polarization state and hence is unpolarized. That is, the Stokes parameters
for each wave must exhibit%
\begin{equation}
\left[ 
\begin{array}{c}
I_{(\mathrm{in})} \\ 
Q_{(\mathrm{in})} \\ 
U_{(\mathrm{in})} \\ 
V_{(\mathrm{in})}%
\end{array}%
\right] \sim \left[ 
\begin{array}{c}
1 \\ 
0 \\ 
0 \\ 
0%
\end{array}%
\right] ,  \label{S}
\end{equation}%
which can be verified via Eqs. (\ref{I0}) - (\ref{V0}) (see Appendix \ref%
{incident}).

\subsubsection{Case 1. For GWs propagating along the line of sight \label%
{alongz}}

As mentioned above, to calculate the EM radiation rescattered by a charge
moving in curved spacetime, we determine the charge's trajectory $X_{\mu
}\left( t\right) $, which is indeed identified with a geodesic worldline
that the charge follows in the spacetime $g_{\mu \nu }=\eta _{\mu \nu
}+h_{\mu \nu }$, where $h_{\mu \nu }$ represents GWs \cite{Ruffini}. For a
weakly perturbed spacetime, in which $\left\vert h_{\mu \nu }\right\vert \ll
1$, one can write down the geodesic equation of motion for the charge in
linearized gravity as follows \cite{Papa}:%
\begin{equation}
\ddot{X}^{i}+\eta ^{ik}h_{jk,t}\dot{X}^{j}+\frac{1}{2}\eta ^{il}\left(
h_{jl,k}+h_{kl,j}-h_{jk,l}\right) \dot{X}^{j}\dot{X}^{k}=\frac{q}{m}\left(
\eta ^{ik}-h_{~}^{ik}\right) E_{k},  \label{geo}
\end{equation}%
where the indices $i$, $j$, $\ldots $ refer to the spatial coordinates $%
\left( x,y,z\right) $, and the overdot `$\cdot $' represents differentiation
with respect to time $t$, and the comma before subscripts indicates partial
differentiation with respect to the subscript that follows the comma. In
this equation the GWs $h_{ij}$ are given by 
\begin{eqnarray}
h_{ij}^{+} &=&h\left( e_{i}^{1}\otimes e_{j}^{1}-e_{i}^{2}\otimes
e_{j}^{2}\right) \exp \left[ \mathrm{i}\omega \left( \frac{z}{c}-t\right) %
\right] ,  \label{hp} \\
h_{ij}^{\times } &=&h\left( e_{i}^{1}\otimes e_{j}^{2}+e_{i}^{2}\otimes
e_{j}^{1}\right) \exp \left[ \mathrm{i}\omega \left( \frac{z}{c}-t\right) %
\right] ,  \label{hc}
\end{eqnarray}%
where $h$ denotes the strain amplitude, $\omega $ the frequency,\footnote{%
In this work, for the sake of simplification, the strain amplitude $h$ is
assumed to be a constant over a finite number of periods $\sim 2n\pi /\omega 
$. Practically, it may be equivalently stated that $h$ changes little for $%
\sim 10^{13}$ seconds: taking into account the redshift effect from
cosmic expansion, the frequency $\omega $ should be $\sim 10^{3}$ times higher than
the characteristic frequency for the regime of the CMB anisotropies in the
spectrum as shown in Figure \ref{fig3}. Later in this Section, the Stokes
parameters are computed for the rescattered EM radiation. They are
time-averaged quantities evaluated over a duration on the same time scale,
which is substantially longer than the characteristic period of the CMB
radiation $\sim 10^{-11}$ seconds.} $e_{i}^{1\left( 2\right) }$ refers to the $x(y)$-component of the unit vector $\mathbf{e}_{i}$, and $+$ and $\times $ represent the two
polarization states prescribed by the tensors $e_{i}^{1}\otimes
e_{j}^{1}-e_{i}^{2}\otimes e_{j}^{2}$ and $e_{i}^{1}\otimes
e_{j}^{2}+e_{i}^{2}\otimes e_{j}^{1}$, respectively. On the right-hand side
of Eq. (\ref{geo}) the EM field is prescribed by means of Eqs. (\ref{Ei})
and (\ref{Eii})) as%
\begin{equation}
E_{i}=e_{i}^{2}E_{\mathrm{I}}\exp \left[ \mathrm{i}\left( \Omega \left( 
\frac{z}{c}-t\right) +\varphi _{\mathrm{I}}\left( t\right) \right) \right]
+e_{i}^{1}E_{\mathrm{II}}\exp \left[ \mathrm{i}\left( \Omega \left( \frac{z}{%
c}-t\right) +\varphi _{\mathrm{II}}\left( t\right) \right) \right] ,
\label{Egeo}
\end{equation}%
which describes the outgoing field reflected by the charge: the charge is
driven to move along the $x$-axis and the $y$-axis by GWs $h_{+}$ or $%
h_{\times }$, and hence reradiates the outgoing reflected light most easily
by moving along the same directions. The incident and outgoing EM radiation
fields and the GWs are illustrated in Figure \ref{fig1}: the GWs propagate
along the line of sight while being polarized in a plane perpendicular to
the line of sight, and the rescattered (outgoing) EM radiation is induced by
the GWs. Note here that the directions of propagation and polarization of
the GWs are fitted into the observer's reference frame; with propagation
along the $z$-axis (coinciding with the line of sight) and polarization
along the $x$-axis and the $y$-axis.

Now, solving the geodesic equation (\ref{geo}) via perturbation, we obtain
the following results (see Appendix \ref{solve} for a detailed derivation):%
\newline
For $h_{+}$,%
\begin{eqnarray}
\dot{x} &=&v_{\mathrm{o}x}-hv_{\mathrm{o}x}\left( 1-\frac{v_{\mathrm{o}z}}{c}%
\right) \exp \left[ \mathrm{i}\omega \left( \frac{z}{c}-t\right) \right] +%
\mathrm{i}\frac{qE_{\mathrm{II}}}{m\Omega }\left\{ \exp \left[ \mathrm{i}%
\left( \Omega \left( \frac{z}{c}-t\right) +\varphi _{\mathrm{II}}\left(
t\right) \right) \right] _{_{_{_{{}}}}}^{^{^{{}}}}\right.  \notag \\
&&\left. -h\left( 1-\frac{v_{\mathrm{o}z}\omega }{c\left( \Omega +\omega
\right) }\right) \exp \left[ \mathrm{i}\left( \left( \Omega +\omega \right)
\left( \frac{z}{c}-t\right) +\varphi _{\mathrm{II}}\left( t\right) \right) %
\right] \right\} +\mathcal{O}_{\mathrm{II}\left[ 0\right] }+\mathcal{O}_{%
\mathrm{II}\left[ h\right] },  \label{px} \\
\dot{y} &=&v_{\mathrm{o}y}+hv_{\mathrm{o}y}\left( 1-\frac{v_{\mathrm{o}z}}{c}%
\right) \exp \left[ \mathrm{i}\omega \left( \frac{z}{c}-t\right) \right] +%
\mathrm{i}\frac{qE_{\mathrm{I}}}{m\Omega }\left\{ \exp \left[ \mathrm{i}%
\left( \Omega \left( \frac{z}{c}-t\right) +\varphi _{\mathrm{I}}\left(
t\right) \right) \right] _{_{_{_{{}}}}}^{^{^{{}}}}\right.  \notag \\
&&\left. +h\left( 1-\frac{v_{\mathrm{o}z}\omega }{c\left( \Omega +\omega
\right) }\right) \exp \left[ \mathrm{i}\left( \left( \Omega +\omega \right)
\left( \frac{z}{c}-t\right) +\varphi _{\mathrm{I}}\left( t\right) \right) %
\right] \right\} +\mathcal{O}_{\mathrm{I}\left[ 0\right] }+\mathcal{O}_{%
\mathrm{I}\left[ h\right] },  \label{py} \\
\dot{z} &=&v_{\mathrm{o}z}-\frac{h\left( v_{\mathrm{o}x}^{2}-v_{\mathrm{o}%
y}^{2}\right) }{2c}\exp \left[ \mathrm{i}\omega \left( \frac{z}{c}-t\right) %
\right]  \notag \\
&&-\mathrm{i}\frac{hq\omega }{mc\Omega \left( \Omega +\omega \right) }%
\left\{ E_{\mathrm{II}}v_{\mathrm{o}x}\exp \left[ \mathrm{i}\left( \left(
\Omega +\omega \right) {\left( \frac{z}{c}-t\right) }+\varphi _{\mathrm{II}%
}\left( t\right) \right) \right] \right.  \notag \\
&&\left. -E_{\mathrm{I}}v_{\mathrm{o}y}\exp \left[ \mathrm{i}\left( \left(
\Omega +\omega \right) {\left( \frac{z}{c}-t\right) }+\varphi _{\mathrm{I}%
}\left( t\right) \right) \right] \right\}  \notag \\
&&+\frac{hq^{2}\omega }{2m^{2}c\Omega ^{2}\left( 2\Omega +\omega \right) }%
\left\{ E_{\mathrm{II}}^{2}\exp \left[ \mathrm{i}\left( \left( 2\Omega
+\omega \right) {\left( \frac{z}{c}-t\right) }+2\varphi _{\mathrm{II}}\left(
t\right) \right) \right] \right.  \notag \\
&&\left. -E_{\mathrm{I}}^{2}\exp \left[ \mathrm{i}\left( \left( 2\Omega
+\omega \right) {\left( \frac{z}{c}-t\right) }+2\varphi _{\mathrm{I}}\left(
t\right) \right) \right] \right\} +\mathcal{O}_{\mathrm{I,II}\left[ 0\right]
}+\mathcal{O}_{\mathrm{I,II}\left[ h\right] },  \label{pz}
\end{eqnarray}%
For $h_{\times }$, 
\begin{eqnarray}
\dot{x} &=&v_{\mathrm{o}x}-hv_{\mathrm{o}y}\left( 1-\frac{v_{\mathrm{o}z}}{c}%
\right) \exp \left[ \mathrm{i}\omega \left( \frac{z}{c}-t\right) \right] +%
\mathrm{i}\frac{qE_{\mathrm{II}}}{m\Omega }\left\{ \exp \left[ \mathrm{i}%
\left( \Omega \left( \frac{z}{c}-t\right) +\varphi _{\mathrm{II}}\left(
t\right) \right) \right] _{_{_{_{{}}}}}^{^{^{{}}}}\right.  \notag \\
&&\left. -h\frac{E_{\mathrm{I}}}{E_{\mathrm{II}}}\left( 1-\frac{v_{\mathrm{o}%
z}\omega }{c\left( \Omega +\omega \right) }\right) \exp \left[ \mathrm{i}%
\left( \left( \Omega +\omega \right) \left( \frac{z}{c}-t\right) +\varphi _{%
\mathrm{I}}\left( t\right) \right) \right] \right\} +\mathcal{O}_{\mathrm{%
I,II}\left[ 0\right] }+\mathcal{O}_{\mathrm{I,II}\left[ h\right] },
\label{cx} \\
\dot{y} &=&v_{\mathrm{o}y}-hv_{\mathrm{o}x}\left( 1-\frac{v_{\mathrm{o}z}}{c}%
\right) \exp \left[ \mathrm{i}\omega \left( \frac{z}{c}-t\right) \right] +%
\mathrm{i}\frac{qE_{\mathrm{I}}}{m\Omega }\left\{ \exp \left[ \mathrm{i}%
\left( \Omega \left( \frac{z}{c}-t\right) +\varphi _{\mathrm{I}}\left(
t\right) \right) \right] _{_{_{_{{}}}}}^{^{^{{}}}}\right.  \notag \\
&&\left. -h\frac{E_{\mathrm{II}}}{E_{\mathrm{I}}}\left( 1-\frac{v_{\mathrm{o}%
z}\omega }{c\left( \Omega +\omega \right) }\right) \exp \left[ \mathrm{i}%
\left( \left( \Omega +\omega \right) \left( \frac{z}{c}-t\right) +\varphi _{%
\mathrm{II}}\left( t\right) \right) \right] \right\} +\mathcal{O}_{\mathrm{%
I,II}\left[ 0\right] }+\mathcal{O}_{\mathrm{I,II}\left[ h\right] },
\label{cy} \\
\dot{z} &=&v_{\mathrm{o}z}-\frac{hv_{\mathrm{o}x}v_{\mathrm{o}y}}{c}\exp %
\left[ \mathrm{i}\omega \left( \frac{z}{c}-t\right) \right]  \notag \\
&&-\mathrm{i}\frac{hq\omega }{mc\Omega \left( \Omega +\omega \right) }%
\left\{ E_{\mathrm{I}}v_{\mathrm{o}x}\exp \left[ \mathrm{i}\left( \left(
\Omega +\omega \right) {\left( \frac{z}{c}-t\right) }+\varphi _{\mathrm{I}%
}\left( t\right) \right) \right] \right.  \notag \\
&&\left. +E_{\mathrm{II}}v_{\mathrm{o}y}\exp \left[ \mathrm{i}\left( \left(
\Omega +\omega \right) {\left( \frac{z}{c}-t\right) }+\varphi _{\mathrm{II}%
}\left( t\right) \right) \right] \right\}  \notag \\
&&+\frac{hq^{2}E_{\mathrm{I}}E_{\mathrm{II}}\omega }{m^{2}c\Omega ^{2}\left(
2\Omega +\omega \right) }\exp \left[ \mathrm{i}\left( \left( 2\Omega +\omega
\right) {\left( \frac{z}{c}-t\right) }+\varphi _{\mathrm{I}}\left( t\right)
+\varphi _{\mathrm{II}}\left( t\right) \right) \right] +\mathcal{O}_{\mathrm{%
I,II}\left[ 0\right] }+\mathcal{O}_{\mathrm{I,II}\left[ h\right] },
\label{cz}
\end{eqnarray}%
where $\left( v_{\mathrm{o}x},v_{\mathrm{o}y},v_{\mathrm{o}z}\right) $
represents the charge's constant drifting velocity, and $\mathcal{O}_{%
\mathrm{I}\left[ 0\right] }\equiv \mathcal{O}\left( \dot{\varphi}_{\mathrm{I}%
}/\Omega ,\ddot{\varphi}_{\mathrm{I}}/\Omega ^{2},\dot{\varphi}_{\mathrm{I}}%
\ddot{\varphi}_{\mathrm{I}}/\Omega ^{3},\dddot{\varphi}_{\mathrm{I}}/\Omega
^{3},\cdots \right) $, $\mathcal{O}_{\mathrm{II}\left[ 0\right] }\equiv 
\mathcal{O}\left( \dot{\varphi}_{\mathrm{II}}/\Omega ,\ddot{\varphi}_{%
\mathrm{II}}/\Omega ^{2},\dot{\varphi}_{\mathrm{II}}\ddot{\varphi}_{\mathrm{%
II}}/\Omega ^{3},\dddot{\varphi}_{\mathrm{II}}/\Omega ^{3},\cdots \right) $, 
$\mathcal{O}_{\mathrm{I,II}\left[ 0\right] }\equiv \mathcal{O}\left( \dot{%
\varphi}_{\mathrm{I}}/\Omega ,\ddot{\varphi}_{\mathrm{I}}/\Omega ^{2},\dot{%
\varphi}_{\mathrm{I}}\ddot{\varphi}_{\mathrm{I}}/\Omega ^{3},\dddot{\varphi}%
_{\mathrm{I}}/\Omega ^{3},\cdots ,\dot{\varphi}_{\mathrm{II}}/\Omega ,\ddot{%
\varphi}_{\mathrm{II}}/\Omega ^{2},\dot{\varphi}_{\mathrm{II}}\ddot{\varphi}%
_{\mathrm{II}}/\Omega ^{3},\dddot{\varphi}_{\mathrm{II}}/\Omega ^{3},\cdots
\right) $, $\mathcal{O}_{\mathrm{I}\left[ h\right] }\sim h\times \mathcal{O}%
_{\mathrm{I}\left[ 0\right] }$, $\mathcal{O}_{\mathrm{II}\left[ h\right]
}\sim h\times \mathcal{O}_{\mathrm{II}\left[ 0\right] }$, $\mathcal{O}_{%
\mathrm{I,II}\left[ h\right] }\sim h\times \mathcal{O}_{\mathrm{I,II}\left[ 0%
\right] }$ are the errors\ generated by the approximate solution for $\dot{X}%
^{i}=\left( \dot{x},\dot{y},\dot{z}\right) $: the subscripts $_{\left[ 0%
\right] }$ and $_{\left[ h\right] }$ denote that the errors are generated
from the unperturbed part and the perturbed part of the solution,
respectively. Above, however, we have assumed that the phase modulation
functions $\varphi _{\mathrm{I}}\left( t\right) $ and $\varphi _{\mathrm{II}%
}\left( t\right) $ vary on a time scale much slower than the period of the
wave, namely, $\left\vert \dot{\varphi}_{\mathrm{I}}\right\vert $, $%
\left\vert \dot{\varphi}_{\mathrm{II}}\right\vert \ll \Omega $ (and further $%
\left\vert \ddot{\varphi}_{\mathrm{I}}\right\vert $, $\left\vert \ddot{%
\varphi}_{\mathrm{II}}\right\vert \ll \Omega ^{2}$, $\left\vert \dddot{%
\varphi}_{\mathrm{I}}\right\vert $, $\left\vert \dddot{\varphi}_{\mathrm{II}%
}\right\vert \ll \Omega ^{3}$, etc.), and therefore all these errors can be
regarded as sufficiently small. A more detailed discussion of the errors is
given in Appendix \ref{solve}.

The rescattered EM potentials are calculated by substituting Eqs. (\ref{px})
-- (\ref{cz}) into Eq. (\ref{A}). They are trivially obtained from%
\begin{eqnarray}
A_{x\,(\mathrm{scat})} &=&\frac{q\dot{x}\left( t_{R}\right) }{r},  \label{Ax}
\\
A_{y\,(\mathrm{scat})} &=&\frac{q\dot{y}\left( t_{R}\right) }{r},  \label{Ay}
\\
A_{z\,(\mathrm{scat})} &=&\frac{q\dot{z}\left( t_{R}\right) }{r},  \label{Az}
\end{eqnarray}%
together with Eqs. (\ref{px}), (\ref{py}), (\ref{pz}) for $h_{+}$ and Eqs. (
\ref{cx}), (\ref{cy}), (\ref{cz}) for $h_{\times }$.

Finally, the rescattered EM fields are calculated by means of $F_{ab}=\nabla
_{a}A_{b}-\nabla _{b}A_{a}$ and Eqs. (\ref{Ax}), (\ref{Ay}), (\ref{Az}).
They are obtained from 
\begin{eqnarray}
E_{x\,(\mathrm{scat})} &=&\partial _{t}A_{x\,(\mathrm{scat})}=\frac{q\ddot{x}%
\left( t_{R}\right) }{r},  \label{Ex} \\
E_{y\,(\mathrm{scat})} &=&\partial _{t}A_{y\,(\mathrm{scat})}=\frac{q\ddot{y}%
\left( t_{R}\right) }{r},  \label{Ey} \\
E_{z\,(\mathrm{scat})} &=&\partial _{t}A_{z\,(\mathrm{scat})}=\frac{q\ddot{z}%
\left( t_{R}\right) }{r},  \label{Ez}
\end{eqnarray}%
together with Eqs. (\ref{px}), (\ref{py}), (\ref{pz}) for $h_{+}$ and Eqs. (%
\ref{cx}), (\ref{cy}), (\ref{cz}) for $h_{\times }$. The results are the
following (see Appendix \ref{solve} for a detailed derivation):\newline
For $h_{+}$,%
\begin{eqnarray}
E_{x\,(\mathrm{scat})} &=&\mathrm{i}\frac{hqv_{\mathrm{o}x}\left( 1-\frac{v_{%
\mathrm{o}z}}{c}\right) \omega }{r}\exp \left[ \mathrm{i}\omega \left( \frac{%
z}{c}-t_{R}\right) \right] +\frac{q^{2}E_{\mathrm{II}}}{mr}\left\{ \exp %
\left[ \mathrm{i}\left( \Omega \left( \frac{z}{c}-t_{R}\right) +\varphi _{%
\mathrm{II}}\left( t_{R}\right) \right) \right] \right.  \notag \\
&&\left. -h\left( 1+\left( 1-\frac{v_{\mathrm{o}z}}{c}\right) \frac{\omega }{%
\Omega }\right) \exp \left[ \mathrm{i}\left( \left( \Omega +\omega \right)
\left( \frac{z}{c}-t_{R}\right) +\varphi _{\mathrm{II}}\left( t_{R}\right)
\right) \right] \right\} +\mathcal{O}_{\mathrm{II}\left[ h\right] },
\label{Epx} \\
E_{y\,(\mathrm{scat})} &=&-\mathrm{i}\frac{hqv_{\mathrm{o}y}\left( 1-\frac{%
v_{\mathrm{o}z}}{c}\right) \omega }{r}\exp \left[ \mathrm{i}\omega \left( 
\frac{z}{c}-t_{R}\right) \right] +\frac{q^{2}E_{\mathrm{I}}}{mr}\left\{ \exp %
\left[ \mathrm{i}\left( \Omega \left( \frac{z}{c}-t_{R}\right) +\varphi _{%
\mathrm{I}}\left( t_{R}\right) \right) \right] \right.  \notag \\
&&\left. +h\left( 1+\left( 1-\frac{v_{\mathrm{o}z}}{c}\right) \frac{\omega }{%
\Omega }\right) \exp \left[ \mathrm{i}\left( \left( \Omega +\omega \right)
\left( \frac{z}{c}-t_{R}\right) +\varphi _{\mathrm{I}}\left( t_{R}\right)
\right) \right] \right\} +\mathcal{O}_{\mathrm{I}\left[ h\right] },
\label{Epy} \\
E_{z\,(\mathrm{scat})} &=&\mathrm{i}\frac{hq\left( v_{\mathrm{o}x}^{2}-v_{%
\mathrm{o}y}^{2}\right) \omega }{2cr}\exp \left[ \mathrm{i}\omega \left( 
\frac{z}{c}-t_{R}\right) \right]  \notag \\
&&-\frac{hq^{2}\omega }{mcr\Omega }\left\{ E_{\mathrm{II}}v_{\mathrm{o}%
x}\exp \left[ \mathrm{i}\left( \left( \Omega +\omega \right) {\left( \frac{z%
}{c}-t_{R}\right) }+\varphi _{\mathrm{II}}\left( t_{R}\right) \right) \right]
\right.  \notag \\
&&\left. -E_{\mathrm{I}}v_{\mathrm{o}y}\exp \left[ \mathrm{i}\left( \left(
\Omega +\omega \right) {\left( \frac{z}{c}-t_{R}\right) }+\varphi _{\mathrm{I%
}}\left( t_{R}\right) \right) \right] \right\}  \notag \\
&&-\mathrm{i}\frac{hq^{3}\omega }{2m^{2}cr\Omega ^{2}}\left\{ E_{\mathrm{II}%
}^{2}\exp \left[ \mathrm{i}\left( \left( 2\Omega +\omega \right) {\left( 
\frac{z}{c}-t_{R}\right) }+2\varphi _{\mathrm{II}}\left( t_{R}\right)
\right) \right] \right.  \notag \\
&&\left. -E_{\mathrm{I}}^{2}\exp \left[ \mathrm{i}\left( \left( 2\Omega
+\omega \right) {\left( \frac{z}{c}-t_{R}\right) }+2\varphi _{\mathrm{I}%
}\left( t_{R}\right) \right) \right] \right\} +\mathcal{O}_{\mathrm{I,II}%
\left[ h\right] },  \label{Epz}
\end{eqnarray}%
For $h_{\times }$,%
\begin{eqnarray}
E_{x\,(\mathrm{scat})} &=&\mathrm{i}\frac{hqv_{\mathrm{o}y}\left( 1-\frac{v_{%
\mathrm{o}z}}{c}\right) \omega }{r}\exp \left[ \mathrm{i}\omega \left( \frac{%
z}{c}-t_{R}\right) \right] +\frac{q^{2}E_{\mathrm{II}}}{mr}\left\{ \exp %
\left[ \mathrm{i}\left( \Omega \left( \frac{z}{c}-t_{R}\right) +\varphi _{%
\mathrm{II}}\left( t_{R}\right) \right) \right] _{_{_{_{{}}}}}^{^{^{{}}}}%
\right.  \notag \\
&&\left. -h\frac{E_{\mathrm{I}}}{E_{\mathrm{II}}}\left( 1+\left( 1-\frac{v_{%
\mathrm{o}z}}{c}\right) \frac{\omega }{\Omega }\right) \exp \left[ \mathrm{i}%
\left( \left( \Omega +\omega \right) \left( \frac{z}{c}-t_{R}\right)
+\varphi _{\mathrm{I}}\left( t_{R}\right) \right) \right] \right\} +\mathcal{%
O}_{\mathrm{I,II}\left[ h\right] },  \label{Ecx} \\
E_{y\,(\mathrm{scat})} &=&\mathrm{i}\frac{hqv_{\mathrm{o}x}\left( 1-\frac{v_{%
\mathrm{o}z}}{c}\right) \omega }{r}\exp \left[ \mathrm{i}\omega \left( \frac{%
z}{c}-t_{R}\right) \right] +\frac{q^{2}E_{\mathrm{I}}}{mr}\left\{ \exp \left[
\mathrm{i}\left( \Omega \left( \frac{z}{c}-t_{R}\right) +\varphi _{\mathrm{I}%
}\left( t_{R}\right) \right) \right] _{_{_{_{{}}}}}^{^{^{{}}}}\right.  \notag
\\
&&\left. -h\frac{E_{\mathrm{II}}}{E_{\mathrm{I}}}\left( 1+\left( 1-\frac{v_{%
\mathrm{o}z}}{c}\right) \frac{\omega }{\Omega }\right) \exp \left[ \mathrm{i}%
\left( \left( \Omega +\omega \right) \left( \frac{z}{c}-t_{R}\right)
+\varphi _{\mathrm{II}}\left( t_{R}\right) \right) \right] \right\} +%
\mathcal{O}_{\mathrm{I,II}\left[ h\right] },  \label{Ecy} \\
E_{z\,(\mathrm{scat})} &=&\mathrm{i}\frac{hqv_{\mathrm{o}x}v_{\mathrm{o}%
y}\omega }{cr}\exp \left[ \mathrm{i}\omega \left( \frac{z}{c}-t_{R}\right) %
\right]  \notag \\
&&-\frac{hq^{2}\omega }{mcr\Omega }\left\{ E_{\mathrm{I}}v_{\mathrm{o}x}\exp %
\left[ \mathrm{i}\left( \left( \Omega +\omega \right) {\left( \frac{z}{c}%
-t_{R}\right) }+\varphi _{\mathrm{I}}\left( t_{R}\right) \right) \right]
\right.  \notag \\
&&\left. +E_{\mathrm{II}}v_{\mathrm{o}y}\exp \left[ \mathrm{i}\left( \left(
\Omega +\omega \right) {\left( \frac{z}{c}-t_{R}\right) }+\varphi _{\mathrm{%
II}}\left( t_{R}\right) \right) \right] \right\}  \notag \\
&&-\mathrm{i}\frac{hq^{3}E_{\mathrm{I}}E_{\mathrm{II}}\omega }{m^{2}cr\Omega
^{2}}\exp \left[ \mathrm{i}\left( \left( 2\Omega +\omega \right) {\left( 
\frac{z}{c}-t_{R}\right) }+\varphi _{\mathrm{I}}\left( t_{R}\right) +\varphi
_{\mathrm{II}}\left( t_{R}\right) \right) \right] +\mathcal{O}_{\mathrm{I,II}%
\left[ h\right] }.  \label{Ecz}
\end{eqnarray}%
Here we note that the errors from the unperturbed part, $\mathcal{O}_{
\mathrm{I}\left[ 0\right] }$, $\mathcal{O}_{\mathrm{II}\left[ 0\right] }$
and $\mathcal{O}_{\mathrm{I,II}\left[ 0\right] }$ are absent and only the
errors from the perturbed part, $\mathcal{O}_{\mathrm{I}\left[ h\right] }$, $
\mathcal{O}_{\mathrm{II}\left[ h\right] }$ and $\mathcal{O}_{\mathrm{I,II}
\left[ h\right] }$ are present. This is because $E_{\,(\mathrm{scat}
)}^{i}\sim \ddot{X}^{i}$ due to Eqs. (\ref{Ex}), (\ref{Ey}) and (\ref{Ez}),
where the unperturbed part of a solution for $\ddot{X}^{i}=\left( \ddot{x},
\ddot{y},\ddot{z}\right) $ is trivially obtained from the right-hand side of
the geodesic equation (\ref{geo}), which is equivalent to Eq. (\ref{Egeo})
apart from the factor; thereby not generating $\mathcal{O}_{\left[ 0\right]
} $. Later in Subsection \ref{Stokes}, we will disregard any contributions
from the errors $\mathcal{O}_{\mathrm{I}\left[ h\right] }$, $\mathcal{O}_{
\mathrm{II}\left[ h\right] }$ and $\mathcal{O}_{\mathrm{I,II}\left[ h\right]
}$ in computing Stokes parameters. A more detailed discussion of the errors
is given in Appendix \ref{solve}.

\subsubsection{Case 2. For GWs propagating along an arbitrary direction 
\label{general}}

Our physical quantities for \textit{Case 1} above were expressed in the
observer's reference frame associated with Cartesian coordinates $\left(
x,y,z\right) $: the rescattered EM radiation was induced by GWs which
propagate along the $z$-axis with polarization along the $x$-axis and the $y$
-axis. 
This does \emph{not} imply that radiation can only be observed along the $z$-axis: we now discuss the case that line of sight and $z$-axis do not coincide.
A new reference frame can be defined by rotating the Cartesian frame
through Euler angles $\left\{ \phi ,\theta ,\psi \right\} $ following the convention by  \cite{Goldstein}:
\begin{equation}
\mathbf{x}^{\prime }=\mathbf{R}\left( \phi ,\theta ,\psi \right) \mathbf{x,}
\label{tr}
\end{equation}%
where $\mathbf{x}^{\prime }\equiv \left( x^{\prime },y^{\prime },z^{\prime
}\right) $ refer to the new frame while $\mathbf{x}\equiv \left(
x,y,z\right) $ the original frame, and%
\begin{eqnarray}
\mathbf{R}\left( \phi ,\theta ,\psi \right) &=&\mathbf{R}_{3}\left( \psi
\right) \mathbf{R}_{2}\left( \theta \right) \mathbf{R}_{1}\left( \phi \right)
\notag \\
&=&\left[ 
\begin{array}{ccc}
\cos \psi \cos \phi -\cos \theta \sin \phi \sin \psi & \cos \psi \sin \phi
+\cos \theta \cos \phi \sin \psi & \sin \psi \sin \theta \\ 
-\sin \psi \cos \phi -\cos \theta \sin \phi \cos \psi & -\sin \psi \sin \phi
+\cos \theta \cos \phi \cos \psi & \cos \psi \sin \theta \\ 
\sin \theta \sin \phi & -\sin \theta \cos \phi & \cos \theta%
\end{array}%
\right] ,  \label{R}
\end{eqnarray}%
with 
\begin{equation}
\mathbf{R}_{1}\left( \phi \right) \equiv \left[ 
\begin{array}{ccc}
\cos \phi & \sin \phi & 0 \\ 
-\sin \phi & \cos \phi & 0 \\ 
0 & 0 & 1%
\end{array}%
\right] ,\mathbf{R}_{2}\left( \theta \right) \equiv \left[ 
\begin{array}{ccc}
1 & 0 & 0 \\ 
0 & \cos \theta & \sin \theta \\ 
0 & -\sin \theta & \cos \theta%
\end{array}%
\right] ,\mathbf{R}_{3}\left( \psi \right) \equiv \left[ 
\begin{array}{ccc}
\cos \psi & \sin \psi & 0 \\ 
-\sin \psi & \cos \psi & 0 \\ 
0 & 0 & 1%
\end{array}%
\right] ,  \label{R1}
\end{equation}%
and $\left\{ \phi ,\theta \right\} $ denote the direction angles in
spherical coordinates, defined with respect to the Cartesian coordinates $%
\left( x,y,z\right) $, and $\psi $ denotes the polarization-ellipse angle 
\cite{Pai}.

\begin{figure}[tbp]
\begin{center}
\includegraphics[scale=0.50]{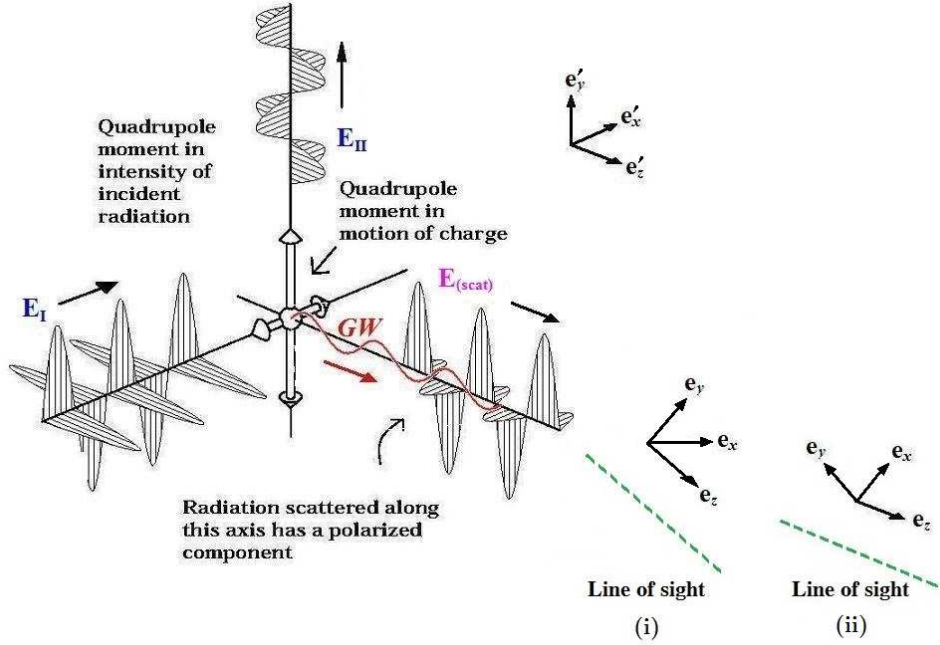}
\end{center}
\caption{An anisotropic incident EM radiation is rescattered by a charge
sitting on spacetime ripples (i.e. GWs) and becomes polarized. Here the
incident and outgoing EM radiation fields and the GWs are illustrated with
respect to the source frame $\left( x^{\prime },y^{\prime },z^{\prime
}\right) $: the GWs propagate along the $z^{\prime }$-axis while being
polarized along the $x^{\prime }$-axis and the $y^{\prime }$-axis, and the
rescattered (outgoing) EM radiation induced by the GWs should be expressed
in the same frame. Concerning the relationship between the source frame $%
\left( x^{\prime },y^{\prime },z^{\prime }\right) $ and the observer's frame 
$\left( x,y,z\right) $, we note the two cases: (i) the propagation direction
of the GWs does not coincide with the line of sight, i.e. $\mathbf{e}%
_{z}^{\prime }$ $\neq \mathbf{e}_{z}$, (ii) the propagation direction of the
GWs coincides with the line of sight, i.e. $\mathbf{e}_{z}^{\prime }$ $=%
\mathbf{e}_{z}$. (Credit: CAPMAP website, 
\texttt{http://quiet.uchicago.edu/capmap}; reproduced with some modifications)
}
\label{fig2}
\end{figure}

Let us now consider a more general situation in which GWs propagate along a
general direction through the CMB. One may let the propagation direction of
our GWs coincide with the $z^{\prime }$-axis in the new frame $\left(
x^{\prime },y^{\prime },z^{\prime }\right) $ which is obtained by rotating
the original frame $\left( x,y,z\right) $ via Eq. (\ref{tr}) above. Then a
charge $q$ in the CMB would be driven to move along the $x^{\prime }$-axis
and the $y^{\prime }$-axis by GWs $h_{+}$ or $h_{\times }$ prescribed in the
new frame. Therefore, the geodesic equation (\ref{geo}), together with Eqs. (%
\ref{hp}), (\ref{hc}) and (\ref{Egeo}) should be rewritten in the
coordinates $\left( t,x^{\prime },y^{\prime },z^{\prime }\right) $. Solving
the geodesic equation in the new frame, one should obtain the same results
as (\ref{px}) - (\ref{cz}) in \textit{Case 1} above, except that the
solutions are now expressed in $\left( t,x^{\prime },y^{\prime },z^{\prime
}\right) $ rather than $\left( t,x,y,z\right) $. The same argument goes for
the rescattered EM fields (\ref{Epx}) - (\ref{Ecz}). In Figure \ref{fig2}
are illustrated the incident and outgoing EM radiation fields and the GWs
with respect to the new coordinate frame, namely the source frame $\left(
x^{\prime },y^{\prime },z^{\prime }\right) $: the GWs propagate along the $%
z^{\prime }$-axis while being polarized along the $x^{\prime }$-axis and the 
$y^{\prime }$-axis, and the rescattered (outgoing) EM radiation induced by
the GWs should be expressed in the same frame. Here, concerning the
relationship between the source frame $\left( x^{\prime },y^{\prime
},z^{\prime }\right) $ and the observer's frame $\left( x,y,z\right) $, we
note the two cases: (i) the propagation direction of the GWs does not
coincide with the line of sight, i.e. $\mathbf{e}_{z}^{\prime }$ $\neq 
\mathbf{e}_{z}$, (ii) the propagation direction of the GWs coincides with
the line of sight, i.e. $\mathbf{e}_{z}^{\prime }$ $=\mathbf{e}_{z}$.

Now, the inverse transformation to Eq. (\ref{tr}), 
\begin{equation}
\mathbf{x}=\mathbf{R}^{-1}\left( \phi ,\theta ,\psi \right) \mathbf{x}%
^{\prime }  \label{tr2}
\end{equation}%
is given by $\mathbf{R}^{\mathrm{T}}\left( \phi ,\theta ,\psi \right) $, the
transpose of the matrix $\mathbf{R}\left( \phi ,\theta ,\psi \right) $ \cite%
{Goldstein}:%
\begin{eqnarray}
\mathbf{R}^{-1}\left( \phi ,\theta ,\psi \right) &=&\mathbf{R}^{\mathrm{T}%
}\left( \phi ,\theta ,\psi \right)  \notag \\
&=&\left[ 
\begin{array}{ccc}
\cos \psi \cos \phi -\cos \theta \sin \phi \sin \psi & -\sin \psi \cos \phi
-\cos \theta \sin \phi \cos \psi & \sin \theta \sin \phi \\ 
\cos \psi \sin \phi +\cos \theta \cos \phi \sin \psi & -\sin \psi \sin \phi
+\cos \theta \cos \phi \cos \psi & -\sin \theta \cos \phi \\ 
\sin \theta \sin \psi & \sin \theta \cos \psi & \cos \theta%
\end{array}%
\right] .  \label{tr3}
\end{eqnarray}%
Using Eq. (\ref{tr2}), the rescattered EM radiation as projected into the
observer's frame $\left( x,y,z\right) $ can be expressed as 
\begin{equation}
\left[ 
\begin{array}{c}
E_{x\,(\mathrm{scat})} \\ 
E_{y\,(\mathrm{scat})} \\ 
E_{z\,(\mathrm{scat})}%
\end{array}%
\right] =\mathbf{R}^{-1}\left( \phi ,\theta ,\psi \right) \left[ 
\begin{array}{c}
E_{x^{\prime }\,(\mathrm{scat})} \\ 
E_{y^{\prime }\,(\mathrm{scat})} \\ 
E_{z^{\prime }\,(\mathrm{scat})}%
\end{array}%
\right] ,  \label{tr4}
\end{equation}%
where the quantity $\left( E_{x^{\prime }\,(\mathrm{scat})},E_{y^{\prime }\,(%
\mathrm{scat})},E_{z^{\prime }\,(\mathrm{scat})}\right) $ is to be taken
from Eqs. (\ref{Epx}) -- (\ref{Ecz}), and the values $\left\{ \phi ,\theta
,\psi \right\} $ are to be specified for the configuration of the source
frame $\left( x^{\prime },y^{\prime },z^{\prime }\right) $ relative to the
observer's frame $\left( x,y,z\right) $. For convenience in Sections \ref%
{Stokes} and \ref{sphere} later, we represent the transformation matrix by%
\begin{equation}
\mathbf{R}^{-1}\left( \phi ,\theta ,\psi \right) =\left[ 
\begin{array}{lll}
a_{11} & a_{12} & a_{13} \\ 
a_{21} & a_{22} & a_{23} \\ 
a_{31} & a_{32} & a_{33}%
\end{array}%
\right] ,  \label{tr5}
\end{equation}%
where $a_{ij}$ ($i,j=1,2,3$) can be directly read off from Eq. (\ref{tr3}).

However, as in the case (i) of Figure \ref{fig2}, when the propagation
direction of the GWs does not coincide with the line of sight, i.e. $\mathbf{
e}_{z}^{\prime }$ $\neq \mathbf{e}_{z}$, the rescattered EM radiation induced by the GWs cannot
actually be detected by an observer who is very far away from the source;
for $r\gg \mathrm{source~size}$ in Eqs. (\ref{Epx}) -- (\ref{Ecz}). That is,
only in the case (ii) of Figure \ref{fig2}, namely, when the propagation
direction of the GWs coincides with the line of sight, i.e. $\mathbf{e}%
_{z}^{\prime }$ $=\mathbf{e}_{z}$, the rescattered EM radiation will be
detectable by a distant observer. This will let us fix the direction angles $%
\left\{ \phi ,\theta \right\} =\left\{ 0,0\right\} $ and reduce $\mathbf{R}%
^{-1}\left( \phi ,\theta ,\psi \right) $ to 
\begin{equation}
\mathbf{R}^{-1}\left( 0,0,\psi \right) =\left[ 
\begin{array}{ccc}
\cos \psi & -\sin \psi & 0 \\ 
\sin \psi & \cos \psi & 0 \\ 
0 & 0 & 1%
\end{array}%
\right] ,  \label{tr6}
\end{equation}%
which simply represents the rotation through the polarization-ellipse angle $%
\psi $.

\subsection{Stokes parameters for the rescattered EM radiation \label{Stokes}%
}

In the previous Subsection we have modeled a situation, in which an
anisotropic incident radiation is rescattered by a charge being agitated by
GWs and the rescattered radiation is indeed induced by the GWs. The charge
is set in motion along particular directions by the GWs $h_{+}$ or $%
h_{\times }$, and this in fact opens \textit{polarization channels} for the
rescattered radiation. For example, a charge being driven to move along the $%
x$-axis and the $y$-axis by GWs $h_{+}$ or $h_{\times }$ will reradiate the
reflected light most easily by moving along the same directions. Now, if the
incident unpolarized radiation is more intense along one axis than the
other, namely, $E_{\mathrm{I}}\neq E_{\mathrm{II}}$ in Eqs. (\ref{Ei}) and (\ref{Eii}), it causes the charge to oscillate more strongly along one axis
than the other, i.e. along the $x$-axis [$y$-axis] than the $y$-axis [$x$%
-axis], for rescattering (see Figures \ref{fig1} and \ref{fig2}). Then the
radiation emitted by this accelerating charge, which propagates along the
third axis, namely, the $z$-axis, has a \textit{net polarization}.
Therefore, the following can be said of the polarization of the CMB induced
by primordial GWs: as primordial GWs agitate a charge in the CMB, setting it
in periodic, regular motion, the EM radiation rescattered by the charge is
partially polarized and this will imprint a characteristic pattern on the
light of the CMB. A net \textit{linear} polarization is expected from the
rescattered radiation, and it can be shown by determining the \textit{Stokes
parameters}. For the two cases of GWs propagating through the CMB, we
present the Stokes parameters as below.

\subsubsection{Case 1. For GWs propagating along the line of sight \label%
{alongz1}}

For the rescattered EM radiation given by (\ref{Epx}) - (\ref{Ecz}), the
Stokes parameters can be calculated in a\ straightforward manner:%
\begin{eqnarray}
I_{(\mathrm{case1})} &=&\left\langle \left\vert E_{x\,(\mathrm{scat}%
)}\right\vert ^{2}\right\rangle +\left\langle \left\vert E_{y\,(\mathrm{scat}%
)}\right\vert ^{2}\right\rangle ,  \label{I0} \\
Q_{(\mathrm{case1})} &=&\left\langle \left\vert E_{x\,(\mathrm{scat}%
)}\right\vert ^{2}\right\rangle -\left\langle \left\vert E_{y\,(\mathrm{scat}%
)}\right\vert ^{2}\right\rangle ,  \label{Q0} \\
U_{(\mathrm{case1})} &=&\left\langle E_{x\,(\mathrm{scat})}E_{y\,(\mathrm{%
scat})}^{\ast }\right\rangle +\left\langle E_{x\,(\mathrm{scat})}^{\ast
}E_{y\,(\mathrm{scat})}\right\rangle ,  \label{U0} \\
V_{(\mathrm{case1})} &=&\mathrm{i}\left( \left\langle E_{x\,(\mathrm{scat}%
)}E_{y\,(\mathrm{scat})}^{\ast }\right\rangle -\left\langle E_{x\,(\mathrm{%
scat})}^{\ast }E_{y\,(\mathrm{scat})}\right\rangle \right) ,  \label{V0}
\end{eqnarray}%
where $\left\langle {}\right\rangle $ represents the time average of the
enclosed quantity and $^{\ast }$ denotes the complex conjugate. After some
calculations, these parameters turn out to exhibit linear polarization as
expected: 
\begin{eqnarray}
I_{(\mathrm{case1})\,\overset{+}{\times }} &=&\frac{q^{4}\left( E_{\mathrm{I}%
}^{2}+E_{\mathrm{II}}^{2}\right) }{m^{2}r^{2}}\left[ 1+h^{2}\left( 1+\left(
1-\frac{v_{\mathrm{o}z}}{c}\right) \frac{\omega }{\Omega }\right) ^{2}\right]
+\frac{h^{2}q^{2}\left( v_{\mathrm{o}x}^{2}+v_{\mathrm{o}y}^{2}\right)
\left( 1-\frac{v_{\mathrm{o}z}}{c}\right) ^{2}\omega ^{2}}{r^{2}},
\label{Is} \\
Q_{(\mathrm{case1})\,\overset{+}{\times }} &=&\frac{q^{4}\left( E_{\mathrm{II%
}}^{2}-E_{\mathrm{I}}^{2}\right) }{m^{2}r^{2}}\left[ 1\pm h^{2}\left(
1+\left( 1-\frac{v_{\mathrm{o}z}}{c}\right) \frac{\omega }{\Omega }\right)
^{2}\right] \pm \frac{h^{2}q^{2}\left( v_{\mathrm{o}x}^{2}-v_{\mathrm{o}%
y}^{2}\right) \left( 1-\frac{v_{\mathrm{o}z}}{c}\right) ^{2}\omega ^{2}}{%
r^{2}},  \label{Qs} \\
U_{(\mathrm{case1})\,\overset{+}{\times }} &=&0,  \label{Us} \\
V_{(\mathrm{case1})\,\overset{+}{\times }} &=&0,  \label{Vs}
\end{eqnarray}
where the labels $+$ and $\times $ denote the two
polarization states of GWs, and the $\pm $ signs on the right-hand side of
Eq. (\ref{Qs}) follow the order of $\overset{+}{\times }$. In these
calculations, we have disregarded any contributions from the errors $%
\mathcal{O}_{\mathrm{I}\left[ h\right] }$, $\mathcal{O}_{\mathrm{II}\left[ h%
\right] }$ and $\mathcal{O}_{\mathrm{I,II}\left[ h\right] }$: they would be $%
\mathcal{O}_{\left[ h\right] }^{2}\sim h^{2}\times \mathcal{O}_{\left[ 0%
\right] }^{2}$ and thus too small to consider.

\begin{figure}[tbp]
\begin{center}
\includegraphics[scale=0.333]{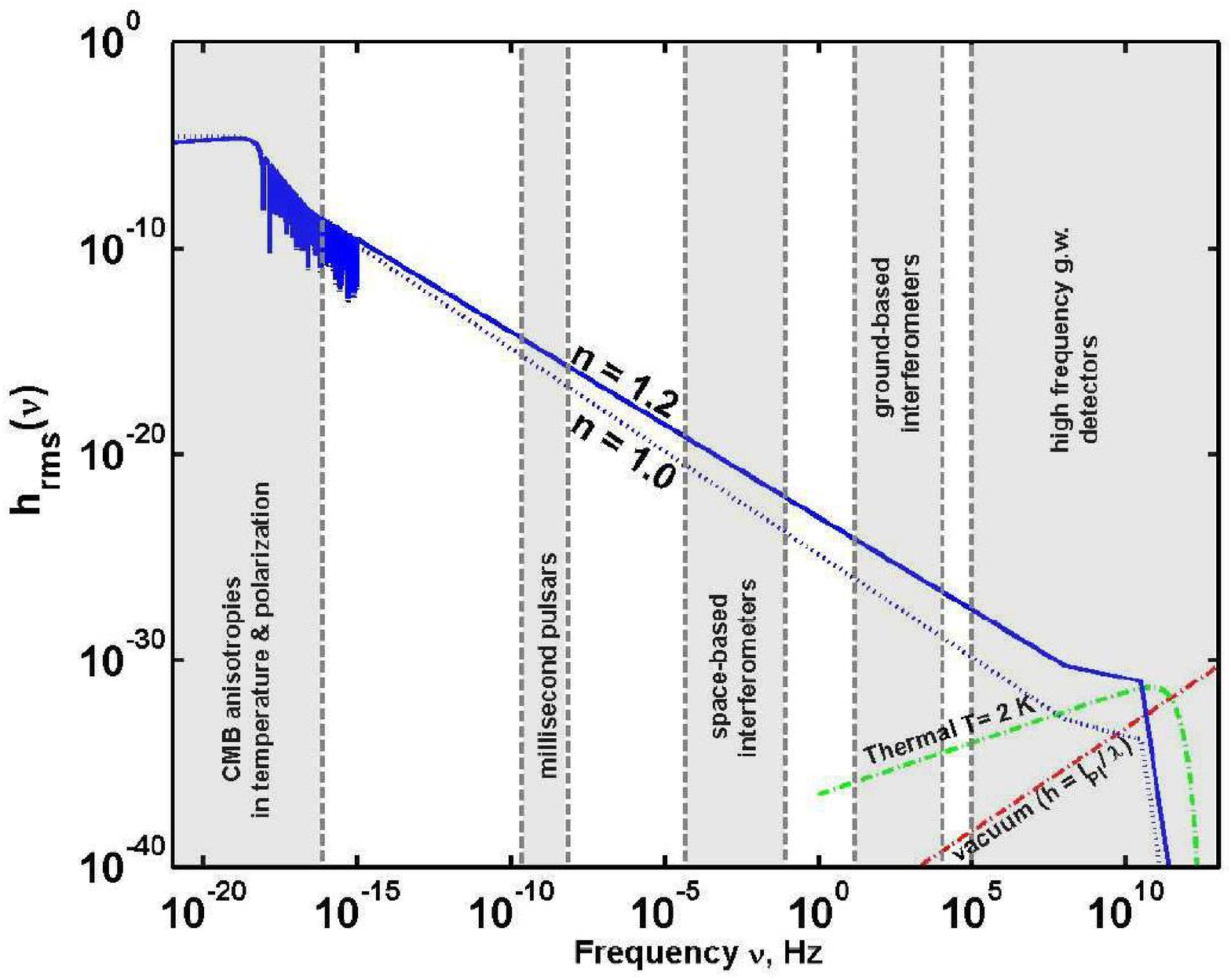}
\end{center}
\caption{The expected spectrum of the characteristic GW amplitude $h$ over a
broad interval of frequencies. (Credit: The Cardiff Gravitational Physics
Group tutorial~\protect\cite{Cardiff})}
\label{fig3}
\end{figure}

In Figure \ref{fig3} is shown the expected spectrum of the characteristic GW
amplitude~over a broad interval of frequencies\ \cite{Cardiff}. The values
of amplitude~in this spectrum have been estimated for present time
observations, and thus our GW amplitude $h$~in Eqs. (\ref{Is}) and (\ref{Qs}%
) should be distinguished from the values of amplitude for the regime of the
CMB anisotropies in the spectrum: our $h$ should be larger than the
characteristic amplitude of primordial GWs $\sim 10^{-10}$ as shown in
Figure \ref{fig3}. Nevertheless, due to the assumption of linearized gravity
as made in Section \ref{EM}, the order of $h^{2}$ should still be regarded
as negligibly small, and hence the terms led by $h^{2}$ in Eqs. (\ref{Is})
and (\ref{Qs}) may be cast off. Taking into consideration the redshift
effect from cosmic expansion, our GW frequency $\omega $ in Eqs. (\ref{Is}) and (%
\ref{Qs}) should be $\sim 10^{3}$ times higher than the characteristic
frequency for the regime of the CMB anisotropies in the spectrum shown in
Figure \ref{fig3}. However, our $\omega $ should still be negligibly small
(in particular, compared to our EM wave frequency $\Omega $), and hence the
terms with $\omega /\Omega $ and $\omega ^{2}$ in Eqs. (\ref{Is}) and (\ref%
{Qs}) may be cast off too. Therefore, no profiles of our GWs would indeed
remain practically in the Stokes parameters (\ref{Is}) - (\ref{Vs}).
Henceforth we may express the parameters simply in the flat spacetime limit: 
\begin{eqnarray}
I_{(\mathrm{case1})\,} &=&\frac{q^{4}\left( E_{\mathrm{I}}^{2}+E_{\mathrm{II}%
}^{2}\right) }{m^{2}r^{2}},  \label{I01} \\
Q_{(\mathrm{case1})\,} &=&\frac{q^{4}\left( E_{\mathrm{II}}^{2}-E_{\mathrm{I}%
}^{2}\right) }{m^{2}r^{2}},  \label{Q01} \\
U_{(\mathrm{case1})\,} &=&0,  \label{U01} \\
V_{(\mathrm{case1})\,} &=&0,  \label{V01}
\end{eqnarray}%
where the expressions are now independent of the polarization states and
therefore the signs $\overset{+}{\times }$ have been dropped.

\subsubsection{Case 2. For GWs propagating along an arbitrary direction 
\label{general1}}

As mentioned in the previous Subsection, our rescattered EM radiation $%
\left( E_{x^{\prime }\,(\mathrm{scat})},E_{y^{\prime }\,(\mathrm{scat}%
)},E_{z^{\prime }\,(\mathrm{scat})}\right) $ in the source frame takes the
same values as Eqs. (\ref{Epx}) -- (\ref{Ecz}), except that it is expressed
in the coordinates $\left( x^{\prime },y^{\prime },z^{\prime }\right) $.
However, to compute the Stokes parameters in the observer's frame $\left(
x,y,z\right) $, the quantity $\left( E_{x^{\prime }\,(\mathrm{scat}%
)},E_{y^{\prime }\,(\mathrm{scat})},E_{z^{\prime }\,(\mathrm{scat})}\right) $
should be projected into that frame (see Figure \ref{fig2} for comparison of
the observer's frame $\left( x,y,z\right) $ and the source frame $\left(
x^{\prime },y^{\prime },z^{\prime }\right) $). Inserting Eqs. (\ref{tr4})
and (\ref{tr5}) into Eqs. (\ref{I0}) -- (\ref{V0}), after rearranging terms,
we have%
\begin{eqnarray}
I_{(\mathrm{case2})} &=&\left( a_{11}^{2}+a_{21}^{2}\right) \left\langle
\left\vert E_{x^{\prime }\,(\mathrm{scat})}\right\vert ^{2}\right\rangle
+\left( a_{12}^{2}+a_{22}^{2}\right) \left\langle \left\vert E_{y^{\prime
}\,(\mathrm{scat})}\right\vert ^{2}\right\rangle   \notag \\
&&+\left( a_{11}a_{12}+a_{21}a_{22}\right) \left( \left\langle E_{x^{\prime
}\,(\mathrm{scat})}E_{y^{\prime }\,(\mathrm{scat})}^{\ast }\right\rangle
+\left\langle E_{x^{\prime }\,(\mathrm{scat})}^{\ast }E_{y^{\prime }\,(%
\mathrm{scat})}\right\rangle \right) ,  \label{I'} \\
Q_{(\mathrm{case2})} &=&\left( a_{11}^{2}-a_{21}^{2}\right) \left\langle
\left\vert E_{x^{\prime }\,(\mathrm{scat})}\right\vert ^{2}\right\rangle
+\left( a_{12}^{2}-a_{22}^{2}\right) \left\langle \left\vert E_{y^{\prime
}\,(\mathrm{scat})}\right\vert ^{2}\right\rangle   \notag \\
&&+\left( a_{11}a_{12}-a_{21}a_{22}\right) \left( \left\langle E_{x^{\prime
}\,(\mathrm{scat})}E_{y^{\prime }\,(\mathrm{scat})}^{\ast }\right\rangle
+\left\langle E_{x^{\prime }\,(\mathrm{scat})}^{\ast }E_{y^{\prime }\,(%
\mathrm{scat})}\right\rangle \right) ,  \label{Q'} \\
U_{(\mathrm{case2})} &=&2a_{11}a_{21}\left\langle \left\vert E_{x^{\prime
}\,(\mathrm{scat})}\right\vert ^{2}\right\rangle +2a_{12}a_{22}\left\langle
\left\vert E_{y^{\prime }\,(\mathrm{scat})}\right\vert ^{2}\right\rangle  
\notag \\
&&+\left( a_{11}a_{22}+a_{12}a_{21}\right) \left( \left\langle E_{x^{\prime
}\,(\mathrm{scat})}E_{y^{\prime }\,(\mathrm{scat})}^{\ast }\right\rangle
+\left\langle E_{x^{\prime }\,(\mathrm{scat})}^{\ast }E_{y^{\prime }\,(%
\mathrm{scat})}\right\rangle \right) ,  \label{U'} \\
V_{(\mathrm{case2})} &=&\mathrm{i}\left( a_{11}a_{22}-a_{12}a_{21}\right)
\left( \left\langle E_{x^{\prime }\,(\mathrm{scat})}E_{y^{\prime }\,(\mathrm{%
scat})}^{\ast }\right\rangle -\left\langle E_{x^{\prime }\,(\mathrm{scat}%
)}^{\ast }E_{y^{\prime }\,(\mathrm{scat})}\right\rangle \right) ,  \label{V'}
\end{eqnarray}%
where $a_{ij}$ is read from Eq. (\ref{tr3}), and any terms associated with $%
E_{z^{\prime }\,(\mathrm{scat})}$ have been disregarded in the flat
spacetime limit: the time average $\left\langle {}\right\rangle $ of these
terms would scale as $h^{2}$, which is negligibly small and
beyond our measurement limit.

Now, from Eqs. (\ref{I0}) -- (\ref{V0}) we may write 
\begin{eqnarray}
\left\langle \left\vert E_{x^{\prime }\,(\mathrm{scat})}\right\vert
^{2}\right\rangle &=&\frac{1}{2}\left( I_{(\mathrm{case1})}^{\prime }+Q_{(%
\mathrm{case1})}^{\prime }\right) ,  \label{I''} \\
\left\langle \left\vert E_{y^{\prime }\,(\mathrm{scat})}\right\vert
^{2}\right\rangle &=&\frac{1}{2}\left( I_{(\mathrm{case1})}^{\prime }-Q_{(%
\mathrm{case1})}^{\prime }\right) ,  \label{Q''} \\
\left\langle E_{x^{\prime }\,(\mathrm{scat})}E_{y^{\prime }\,(\mathrm{scat}%
)}^{\ast }\right\rangle &=&\frac{1}{2}\left( U_{(\mathrm{case1})}^{\prime }-%
\mathrm{i}V_{(\mathrm{case1})}^{\prime }\right) ,  \label{U''} \\
\left\langle E_{x^{\prime }\,(\mathrm{scat})}^{\ast }E_{y^{\prime }\,(%
\mathrm{scat})}\right\rangle &=&\frac{1}{2}\left( U_{(\mathrm{case1}%
)}^{\prime }+\mathrm{i}V_{(\mathrm{case1})}^{\prime }\right) ,  \label{V''}
\end{eqnarray}%
where the Stokes parameters are defined in the source frame $\left(
x^{\prime },y^{\prime },z^{\prime }\right) $: however, their values are
equal to those in Eqs. (\ref{I01}) -- (\ref{V01}) for \textit{Case 1}.
Substituting Eqs. (\ref{I''}) -- (\ref{V''}) into Eqs. (\ref{I'}) -- (\ref{V'}%
), we establish the following relation: 
\begin{equation}
\left[ 
\begin{array}{l}
I_{(\mathrm{case2})} \\ 
Q_{(\mathrm{case2})} \\ 
U_{(\mathrm{case2})} \\ 
V_{(\mathrm{case2})}%
\end{array}%
\right] =\mathbf{L}\left[ 
\begin{array}{l}
I_{(\mathrm{case1})}^{\prime } \\ 
Q_{(\mathrm{case1})}^{\prime } \\ 
U_{(\mathrm{case1})}^{\prime } \\ 
V_{(\mathrm{case1})}^{\prime }%
\end{array}%
\right] ,  \label{SP}
\end{equation}%
where 
\begin{equation}
\mathbf{L=}\frac{1}{2}\left[ 
\begin{array}{cccc}
a_{11}^{2}+a_{12}^{2}+a_{21}^{2}+a_{22}^{2} & 
a_{11}^{2}-a_{12}^{2}+a_{21}^{2}-a_{22}^{2} & 2\left(
a_{11}a_{12}+a_{21}a_{22}\right) & 0 \\ 
a_{11}^{2}+a_{12}^{2}-a_{21}^{2}-a_{22}^{2} & 
a_{11}^{2}-a_{12}^{2}-a_{21}^{2}+a_{22}^{2} & 2\left(
a_{11}a_{12}-a_{21}a_{22}\right) & 0 \\ 
2\left( a_{11}a_{21}+a_{12}a_{22}\right) & 2\left(
a_{11}a_{21}-a_{12}a_{22}\right) & 2\left( a_{11}a_{22}+a_{12}a_{21}\right)
& 0 \\ 
0 & 0 & 0 & 2\left( a_{11}a_{22}-a_{12}a_{21}\right)%
\end{array}%
\right] ,  \label{SP1}
\end{equation}%
with the values $\left\{ I_{(\mathrm{case1})}^{\prime },Q_{(\mathrm{case1}%
)}^{\prime },U_{(\mathrm{case1})}^{\prime },V_{(\mathrm{case1})}^{\prime
}\right\} $ to be taken from Eqs. (\ref{I01}) -- (\ref{V01}). By means of
Eqs. (\ref{tr3}) and (\ref{tr5}) we can write out Eq. (\ref{SP1}):%
\begin{equation}
\mathbf{L=}\frac{1}{2}\left[ 
\begin{array}{cccc}
1+\cos ^{2}\theta & \sin ^{2}\theta \cos \left( 2\psi \right) & -\sin
^{2}\theta \sin \left( 2\psi \right) & 0 \\ 
\sin ^{2}\theta \cos \left( 2\phi \right) & 
\begin{array}{c}
-2\cos \theta \sin \left( 2\phi \right) \sin \left( 2\psi \right) \\ 
\left( 1+\cos ^{2}\theta \right) \cos \left( 2\phi \right) \cos \left( 2\psi
\right)%
\end{array}
& 
\begin{array}{c}
-2\cos \theta \sin \left( 2\phi \right) \cos \left( 2\psi \right) \\ 
-\left( 1+\cos ^{2}\theta \right) \cos \left( 2\phi \right) \sin \left(
2\psi \right)%
\end{array}
& 0 \\ 
\sin ^{2}\theta \sin \left( 2\phi \right) & 
\begin{array}{c}
2\cos \theta \cos \left( 2\phi \right) \sin \left( 2\psi \right) \\ 
+\left( 1+\cos ^{2}\theta \right) \sin \left( 2\phi \right) \cos \left(
2\psi \right)%
\end{array}
& 
\begin{array}{c}
2\cos \theta \cos \left( 2\phi \right) \cos \left( 2\psi \right) \\ 
-\left( 1+\cos ^{2}\theta \right) \sin \left( 2\phi \right) \sin \left(
2\psi \right)%
\end{array}
& 0 \\ 
0 & 0 & 0 & 2\cos \theta%
\end{array}%
\right] .  \label{L}
\end{equation}

By Eqs. (\ref{SP}) and (\ref{L}) together with (\ref{I01}) -- (\ref{V01}), we
can write down explicitly, 
\begin{eqnarray}
I_{(\mathrm{case2})} &=&\frac{1+\cos ^{2}\theta }{2}\frac{q^{4}\left( E_{%
\mathrm{I}}^{2}+E_{\mathrm{II}}^{2}\right) }{m^{2}r^{2}}+\frac{\sin
^{2}\theta \cos \left( 2\psi \right) }{2}\frac{q^{4}\left( E_{\mathrm{II}%
}^{2}-E_{\mathrm{I}}^{2}\right) }{m^{2}r^{2}},  \label{I02} \\
Q_{(\mathrm{case2})} &=&\frac{\sin ^{2}\theta \cos \left( 2\phi \right) }{2}%
\frac{q^{4}\left( E_{\mathrm{I}}^{2}+E_{\mathrm{II}}^{2}\right) }{m^{2}r^{2}}
\notag \\
&&+\left[ -\cos \theta \sin \left( 2\phi \right) \sin \left( 2\psi \right) +%
\frac{\left( 1+\cos ^{2}\theta \right) \cos \left( 2\phi \right) \cos \left(
2\psi \right) }{2}\right] \frac{q^{4}\left( E_{\mathrm{II}}^{2}-E_{\mathrm{I}%
}^{2}\right) }{m^{2}r^{2}}  \label{Q02} \\
U_{(\mathrm{case2})} &=&\frac{\sin ^{2}\theta \sin \left( 2\phi \right) }{2}%
\frac{q^{4}\left( E_{\mathrm{I}}^{2}+E_{\mathrm{II}}^{2}\right) }{m^{2}r^{2}}
\notag \\
&&+\left[ \cos \theta \cos \left( 2\phi \right) \sin \left( 2\psi \right) +%
\frac{\left( 1+\cos ^{2}\theta \right) \sin \left( 2\phi \right) \cos \left(
2\psi \right) }{2}\right] \frac{q^{4}\left( E_{\mathrm{II}}^{2}-E_{\mathrm{I}%
}^{2}\right) }{m^{2}r^{2}},  \label{U02} \\
V_{(\mathrm{case2})} &=&0.  \label{V02}
\end{eqnarray}%
These represent the Stokes parameters as computed by an observer as the
rescattered EM radiation $\left( E_{x^{\prime }\,(\mathrm{scat}
)},E_{y^{\prime }\,(\mathrm{scat})},E_{z^{\prime }\,(\mathrm{scat})}\right) $
in the source frame $\left( x^{\prime },y^{\prime },z^{\prime }\right) $ is
projected into the observer's frame $\left( x,y,z\right) $.

However, the argument presented at the end of Section \ref{general} suggests
that the only physically meaningful transformation from \textit{Case 1} to 
\textit{Case 2} would be the rotation through the polarization-ellipse angle 
$\psi $ while fixing the direction angles $\left\{ \phi ,\theta \right\}
=\left\{ 0,0\right\} $; namely, keeping the propagation direction of GWs
aligned with the line of sight, i.e. $\mathbf{e}_{z}^{\prime }$ $=\mathbf{e}
_{z}$ (see Figure \ref{fig2}). Then Eqs. (\ref{I02}) -- (\ref{V02}) will
reduce to 
\begin{eqnarray}
I_{(\mathrm{case2})} &=&\frac{q^{4}\left( E_{\mathrm{I}}^{2}+E_{\mathrm{II}%
}^{2}\right) }{m^{2}r^{2}},  \label{i0} \\
Q_{(\mathrm{case2})} &=&\cos \left( 2\psi \right) \frac{q^{4}\left( E_{%
\mathrm{II}}^{2}-E_{\mathrm{I}}^{2}\right) }{m^{2}r^{2}},  \label{q0} \\
U_{(\mathrm{case2})} &=&\sin \left( 2\psi \right) \frac{q^{4}\left( E_{%
\mathrm{II}}^{2}-E_{\mathrm{I}}^{2}\right) }{m^{2}r^{2}}  \label{u0} \\
V_{(\mathrm{case2})} &=&0.  \label{v0}
\end{eqnarray}%
In comparison to Eqs. (\ref{I01}) -- (\ref{V01}) for \textit{Case 1}, this
implies that we have the linear polarization components along the axes
rotated from the $x$-axis and the $y$-axis by $\psi $.

\section{CMB polarization observables \label{obs}}

\subsection{The Stokes parameters redefined on a sphere and polarization
patterns \label{sphere}}

In the previous Section we have computed the Stokes parameters for the EM
radiation rescattered by a charge being shaken by GWs propagating in the
CMB. As shown by Eqs. (\ref{I01}) -- (\ref{V01}) and (\ref{i0}) -- (\ref{v0}),
the Stokes parameters exhibit a net linear polarization as desired. However,
to analyze the CMB polarization on the sky, it is natural to redefine the
Stokes parameters on a sphere such that they can be decomposed into
spherical harmonics; to be precise, spin-weighted spherical harmonics as
shall be seen later. Below we describe how this is achieved.

First, recall from the previous Section that the Stokes parameters (\ref{I01}%
) - (\ref{V01}) and (\ref{i0}) - (\ref{v0}) have been computed with respect
to $\left( \mathbf{e}_{x},\mathbf{e}_{y}\right) $ as the rescattered EM
radiation propagates along $\mathbf{e}_{z}$ in the Cartesian coordinate
system $\left( x,y,z\right) $. Now, to redefine the Stokes parameters on a
sphere, one should measure them relative to $\left( \mathbf{e}_{\theta },
\mathbf{e}_{\phi }\right) $ as the rescattered EM radiation propagates along 
$\mathbf{e}_{r}$ in the spherical coordinate system $\left( r,\theta ,\phi
\right) $. This can be achieved via Eq. (\ref{tr}); by rotating the
Cartesian frame through the Euler angles $\left\{ \phi ,\theta \right\} $
(direction angles) while fixing the Euler angle $\psi =0$
(polarization-ellipse angle). This will render the inverse transformation to
Eq. (\ref{tr4}), which can be equivalently obtained by switching $
a_{ij}\leftrightarrow a_{ji}$ (i.e. transpose) and fixing $\psi =0$ in Eqs. (\ref{tr3}) and (\ref{tr5}). Further, we find that the inverse transformation
to (\ref{SP}) can be obtained by applying the same argument to Eq. (\ref{SP1}%
). That is, 
\begin{equation}
\left[ 
\begin{array}{l}
I^{\prime } \\ 
Q^{\prime } \\ 
U^{\prime } \\ 
V^{\prime }%
\end{array}%
\right] =\mathbf{M}\left[ 
\begin{array}{l}
I \\ 
Q \\ 
U \\ 
V%
\end{array}%
\right] ,  \label{SP2}
\end{equation}%
where 
\begin{eqnarray}
\mathbf{M} &\mathbf{=L}&\left( a_{ij}\leftrightarrow a_{ji};\psi =0\right) 
\notag \\
&=&\frac{1}{2}\left[ 
\begin{array}{cccc}
a_{11}^{2}+a_{12}^{2}+a_{21}^{2}+a_{22}^{2} & 
a_{11}^{2}+a_{12}^{2}-a_{21}^{2}-a_{22}^{2} & 2\left(
a_{11}a_{21}+a_{12}a_{22}\right) & 0 \\ 
a_{11}^{2}-a_{12}^{2}+a_{21}^{2}-a_{22}^{2} & 
a_{11}^{2}-a_{12}^{2}-a_{21}^{2}+a_{22}^{2} & 2\left(
a_{11}a_{21}-a_{12}a_{22}\right) & 0 \\ 
2\left( a_{11}a_{12}+a_{21}a_{22}\right) & 2\left(
a_{11}a_{12}-a_{21}a_{22}\right) & 2\left( a_{11}a_{22}+a_{12}a_{21}\right)
& 0 \\ 
0 & 0 & 0 & 2\left( a_{11}a_{22}-a_{12}a_{21}\right)%
\end{array}%
\right] _{\psi =0}  \notag \\
&=&\frac{1}{2}\left[ 
\begin{array}{cccc}
1+\cos ^{2}\theta & \sin ^{2}\theta \cos \left( 2\phi \right) & \sin
^{2}\theta \sin \left( 2\phi \right) & 0 \\ 
\sin ^{2}\theta & \left( 1+\cos ^{2}\theta \right) \cos \left( 2\phi \right)
& \left( 1+\cos ^{2}\theta \right) \sin \left( 2\phi \right) & 0 \\ 
0 & -2\cos \theta \sin \left( 2\phi \right) & 2\cos \theta \cos \left( 2\phi
\right) & 0 \\ 
0 & 0 & 0 & 2\cos \theta%
\end{array}%
\right] ,  \label{SP3}
\end{eqnarray}%
with $a_{ij}$ identified from Eqs. (\ref{tr3}) and (\ref{tr5}). Here Eq. (
\ref{SP2}) means that the Stokes parameters $\left\{ I^{\prime },Q^{\prime
},U^{\prime },V^{\prime }\right\} $ (in the spherical coordinate system) are
equal to the values of the Stokes parameters $\left\{ I,Q,U,V\right\} $ (in
the Cartesian coordinate system) as measured relative to $\left( \mathbf{e}
_{\theta },\mathbf{e}_{\phi }\right) $: the Stokes parameters $\left\{
I^{\prime },Q^{\prime },U^{\prime },V^{\prime }\right\} $ are functions of
spherical polar angles $\left\{ \phi ,\theta \right\} $ while the Stokes
parameters $\left\{ I,Q,U,V\right\} $ are constant evaluated values in the
Cartesian frame, given by Eqs. (\ref{I01}) -- (\ref{V01}) or (\ref{i0}) -- (
\ref{v0}). Hereafter, for notational convenience, we drop the sign $^{\prime
}$ from $\left\{ I^{\prime },Q^{\prime },U^{\prime },V^{\prime }\right\} $
on the left-hand side of Eq. (\ref{SP2}) but put the subscript $_{\mathrm{c}
} $ (to stand for `constant') in $\left\{ I,Q,U,V\right\} $ on the
right-hand side. Then we may rewrite Eq. (\ref{SP2}) as 
\begin{equation}
\left[ 
\begin{array}{l}
I \\ 
Q \\ 
U \\ 
V%
\end{array}%
\right] =\frac{3}{4}\left[ 
\begin{array}{cccc}
1+\cos ^{2}\theta & \sin ^{2}\theta \cos \left( 2\phi \right) & \sin
^{2}\theta \sin \left( 2\phi \right) & 0 \\ 
\sin ^{2}\theta & \left( 1+\cos ^{2}\theta \right) \cos \left( 2\phi \right)
& \left( 1+\cos ^{2}\theta \right) \sin \left( 2\phi \right) & 0 \\ 
0 & -2\cos \theta \sin \left( 2\phi \right) & 2\cos \theta \cos \left( 2\phi
\right) & 0 \\ 
0 & 0 & 0 & 2\cos \theta%
\end{array}%
\right] \left[ 
\begin{array}{l}
I_{\mathrm{c}} \\ 
Q_{\mathrm{c}} \\ 
U_{\mathrm{c}} \\ 
V_{\mathrm{c}}%
\end{array}%
\right] ,  \label{SP4}
\end{equation}%
where $\left\{ I_{\mathrm{c}},Q_{\mathrm{c}},U_{\mathrm{c}},V_{\mathrm{c}
}\right\} $ refer to Eqs. (\ref{I01}) -- (\ref{V01}) or (\ref{i0}) -- (\ref{v0}
), and the factor $1/2$ from the matrix (\ref{SP3}) has been readjusted to $
3/4$ for normalization \cite{Chandra, Halonen}; such that the scattered
intensity integrated over all directions, $0\leq \phi <2\pi $, $0\leq \theta
<\pi $ equal the intensity of the incoming radiation.

To study the linear polarization of the CMB, we examine only $Q$ and $U$
among the Stokes parameters. From Eq. (\ref{SP4}) the combinations of the
two parameters turn out to be 
\begin{eqnarray}
Q\pm \mathrm{i}U &=&\frac{3}{4}I_{\mathrm{c}}\sin ^{2}\theta +Q_{\mathrm{c}}%
\left[ \frac{3}{4}\left( 1+\cos ^{2}\theta \right) \cos \left( 2\phi \right)
\mp \mathrm{i}\frac{3}{2}\cos \theta \sin \left( 2\phi \right) \right] 
\notag \\
&&+U_{\mathrm{c}}\left[ \frac{3}{4}\left( 1+\cos ^{2}\theta \right) \sin
\left( 2\phi \right) \pm \mathrm{i}\frac{3}{2}\cos \theta \cos \left( 2\phi
\right) \right] .  \label{QpmU0}
\end{eqnarray}%
Or we can put this in the representation, 
\begin{equation}
Q\pm \mathrm{i}U=\sqrt{\frac{6\pi }{5}}I_{\mathrm{c}}{}~_{\pm
2}Y_{2}^{0}\left( \theta ,\phi \right) +\sqrt{\frac{9\pi }{5}}\left( Q_{%
\mathrm{c}}-\mathrm{i}U_{\mathrm{c}}\right) ~_{\pm 2}Y_{2}^{2}\left( \theta
,\phi \right) +\sqrt{\frac{9\pi }{5}}\left( Q_{\mathrm{c}}+\mathrm{i}U_{%
\mathrm{c}}\right) ~_{\pm 2}Y_{2}^{-2}\left( \theta ,\phi \right) ,
\label{QpmU}
\end{equation}%
where $_{\pm s}Y_{\ell }^{m}$ represent the spin-weighted spherical
harmonics of spin $\pm s$ \cite{Newman}. This clearly shows that $Q\pm 
\mathrm{i}U$ are spin $\pm 2$ quantities.

However, from Eqs. (\ref{SP}) and (\ref{L}) we can express $\left\{ I_{%
\mathrm{c}},Q_{\mathrm{c}},U_{\mathrm{c}},V_{\mathrm{c}}\right\} $ in the
most general case for observation: 
\begin{eqnarray}
I_{\mathrm{c}} &=&I_{\mathrm{o}\,},  \label{I03} \\
Q_{\mathrm{c}} &=&\cos \left( 2\psi _{\mathrm{o}}\right) Q_{\mathrm{o}%
\,}-\sin \left( 2\psi _{\mathrm{o}}\right) U_{\mathrm{o}},  \label{Q03} \\
U_{\mathrm{c}} &=&\sin \left( 2\psi _{\mathrm{o}}\right) Q_{\mathrm{o}%
\,}+\cos \left( 2\psi _{\mathrm{o}}\right) U_{\mathrm{o}},  \label{U03} \\
V_{\mathrm{c}} &=&V_{\mathrm{o}},  \label{V03}
\end{eqnarray}%
where $\psi _{\mathrm{o}}$ denotes the value of polarization-ellipse angle,
by which the source frame $\left( x^{\prime },y^{\prime },z^{\prime }\right) 
$ is rotated from the observer's frame $\left( x,y,z\right) $ with respect
to the $z$-axis while keeping $\mathbf{e}_{z}^{\prime }$ $=\mathbf{e}_{z}$
(see Figure \ref{fig2} for comparison of the two frames), and 
\begin{eqnarray}
I_{\mathrm{o}\,} &\equiv &I_{(\mathrm{case1})\,}=\frac{q^{4}\left( E_{%
\mathrm{I}}^{2}+E_{\mathrm{II}}^{2}\right) }{m^{2}r^{2}},  \label{I04} \\
Q_{\mathrm{o}\,} &\equiv &Q_{(\mathrm{case1})\,}=\frac{q^{4}\left( E_{%
\mathrm{II}}^{2}-E_{\mathrm{I}}^{2}\right) }{m^{2}r^{2}},  \label{Q04} \\
U_{\mathrm{o}} &\equiv &U_{(\mathrm{case1})\,}=0,  \label{U04} \\
V_{\mathrm{o}} &\equiv &V_{(\mathrm{case1})\,}=0,  \label{V04}
\end{eqnarray}%
by Eqs. (\ref{I01}) -- (\ref{V01}). In the limit $\psi _{\mathrm{o}
}\rightarrow 0$, we have $\left\{ I_{\mathrm{c}},Q_{\mathrm{c}},U_{\mathrm{c}
},V_{\mathrm{c}}\right\} \rightarrow \left\{ I_{\mathrm{o}\,},Q_{\mathrm{o}
\,},U_{\mathrm{o}},V_{\mathrm{o}}\right\} $. Inserting Eqs. (\ref{I03}) -- (
\ref{V03}) into Eq. (\ref{QpmU0}), we may rewrite 
\begin{eqnarray}
Q\pm \mathrm{i}U &=&\frac{3}{4}I_{\mathrm{o}\,}\sin ^{2}\theta   \notag \\
&&+\frac{3}{4}Q_{\mathrm{o}\,}\left[ \left( 1+\cos ^{2}\theta \right) \cos
\left( 2\left( \phi -\psi _{\mathrm{o}}\right) \right) \mp \mathrm{i}2\cos
\theta \sin \left( 2\left( \phi -\psi _{\mathrm{o}}\right) \right) \right] .
\label{QpmU1}
\end{eqnarray}%
Here the term $\frac{3}{4}I_{\mathrm{o}\,}\sin ^{2}\theta $ refers to the
total intensity of radiation and does not concern polarization. Therefore,
we may remove this and redefine%
\begin{equation}
\bar{Q}\pm \mathrm{i}\bar{U}=\frac{3}{4}Q_{\mathrm{o}\,}\left[ \left( 1+\cos
^{2}\theta \right) \cos \left( 2\left( \phi -\psi _{\mathrm{o}}\right)
\right) \mp \mathrm{i}2\cos \theta \sin \left( 2\left( \phi -\psi _{\mathrm{o%
}}\right) \right) \right] .  \label{QpmU2}
\end{equation}%
Or we can represent this using the spin-weighted spherical harmonics,%
\begin{equation}
\bar{Q}\pm \mathrm{i}\bar{U}=\sqrt{\frac{9\pi }{5}}Q_{\mathrm{o}\,}\left[ \
_{\pm 2}Y_{2}^{2}\left( \theta ,\phi -\psi _{\mathrm{o}}\right) +\ _{\pm
2}Y_{2}^{-2}\left( \theta ,\phi -\psi _{\mathrm{o}}\right) \right] .
\label{QpmU3}
\end{equation}

By means of Eqs. (\ref{QpmU2}) or (\ref{QpmU3}), we can illustrate
polarization patterns; namely, \textquotedblleft electric\textquotedblright\
E-mode and \textquotedblleft magnetic\textquotedblright\ B-mode patterns 
\cite{Hu}. The polarization amplitude and angle clockwise from the north
pole (at $\theta =0$) are defined to represent E-modes, and B-modes can be
represented by rotating the E-modes by 45$^{\circ }$:%
\begin{eqnarray}
\text{\textrm{Amplitude}} &=&\sqrt{\bar{Q}^{2}+\bar{U}^{2}}  \notag \\
&=&\frac{3}{4}\left\vert Q_{\mathrm{o}\,}\right\vert \left\{ \left[ \left(
1+\cos ^{2}\theta \right) \cos \left( 2\left( \phi -\psi _{\mathrm{o}%
}\right) \right) \right] ^{2}+\left[ -2\cos \theta \sin \left( 2\left( \phi
-\psi _{\mathrm{o}}\right) \right) \right] ^{2}\right\} ^{1/2},  \label{amp}
\end{eqnarray}%
\begin{eqnarray}
\mathrm{Angle}\left( \text{\textrm{E-modes}}\right) &=&\frac{1}{2}\arctan
\left( \frac{\bar{U}}{\bar{Q}}\right)  \notag \\
&=&\frac{1}{2}\arctan \left( \frac{-2\cos \theta \tan \left( 2\left( \phi
-\psi _{\mathrm{o}}\right) \right) }{1+\cos ^{2}\theta }\right) ,
\label{angE} \\
\mathrm{Angle}\left( \text{\textrm{B-modes}}\right) &=&\mathrm{Angle}\left( 
\text{\textrm{E-modes}}\right) +\frac{\pi }{4},  \label{angB}
\end{eqnarray}%
with $Q_{\mathrm{o}\,}$ given by Eq. (\ref{Q04}). Regarding the definitions of E-modes/B-modes, we followed \cite{Hu}. Eqs. (\ref{angE}) and (\ref{angB}) refer to the angles for E-modes/B-modes derived from the `spin $\pm 2$' quantity as given by Eq. (\ref{QpmU3}); this should be distinguished from the later case of Eqs. (\ref{angE1}) and (\ref{angB1}) which refer to the angles for E-modes/B-modes derived from the `spin $0$' quantity as given by Eq. (\ref{EpmB}).

\begin{center}
\textbf{Graphic representation 1}
\end{center}

\begin{figure}[tbp]
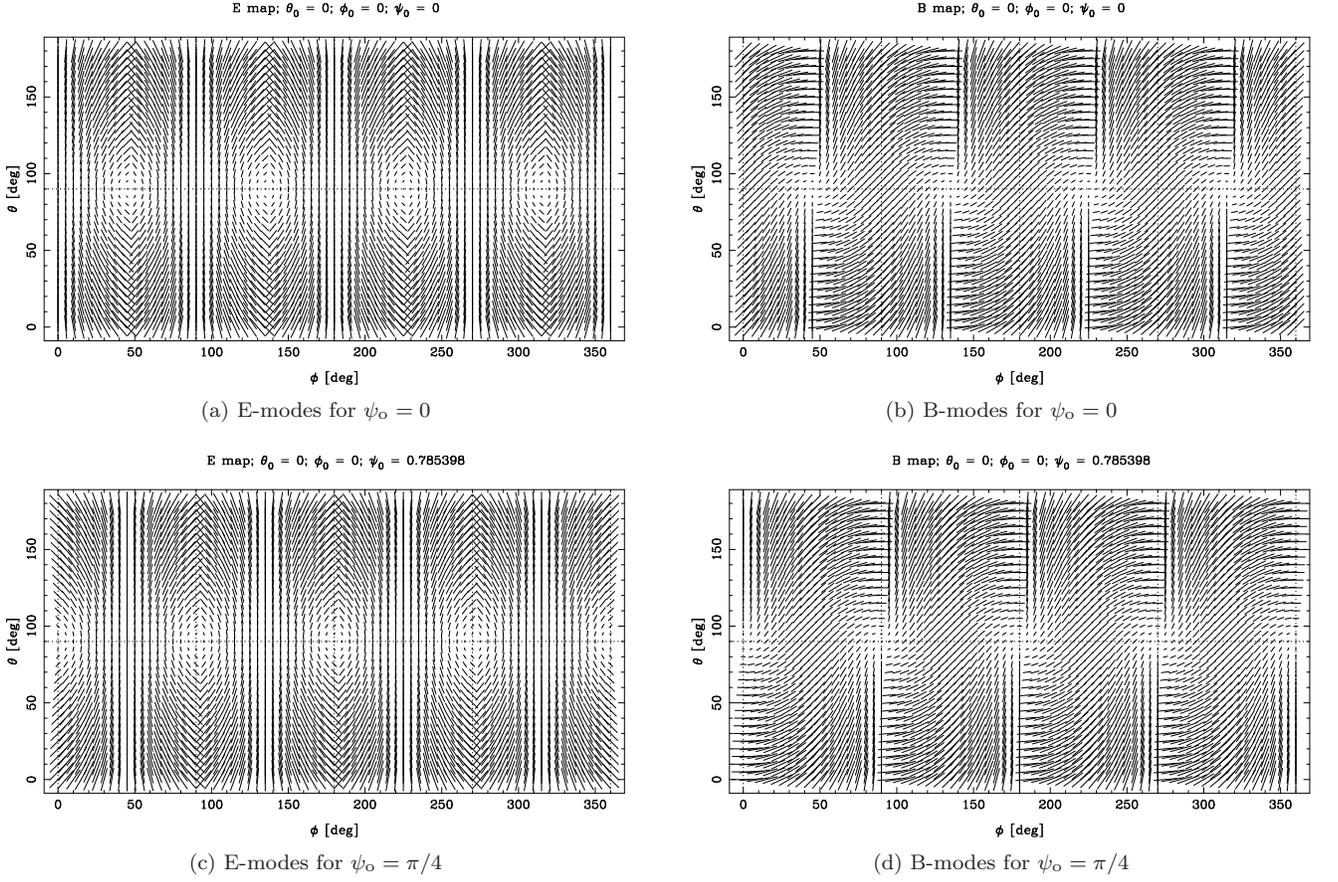

\centering
\subfloat[Subfigure 1 list of figures text][E-modes for $\psi _{\mathrm{o}}=0$]{
\includegraphics[angle=270,width=0.46\textwidth]{E2-map-001}
\label{fig:subfig1}} 
\qquad 
\subfloat[Subfigure 2 list of figures text][B-modes for $\psi _{\mathrm{o}}=0$]{
\includegraphics[angle=270,width=0.46\textwidth]{B2-map-001}
\label{fig:subfig2}} \ \ \ \ \ \ 
\smallskip
\subfloat[Subfigure 3 list of figures text][E-modes for $\psi _{\mathrm{o}}=\pi /4$]{
\includegraphics[angle=270,width=0.46\textwidth]{E2-map-002}
\label{fig:subfig3}} 
\qquad 
\subfloat[Subfigure 4 list of figures text][B-modes for $\psi _{\mathrm{o}}=\pi /4$]{
\includegraphics[angle=270,width=0.46\textwidth]{B2-map-002}
\label{fig:subfig4}} \ \ \ \ \ \ 
\smallskip
\caption{The E-mode and B-mode patterns of polarization given the polarization-ellipse angle $\psi _{\mathrm{o}}\
$.}
\label{fig4}
\end{figure}

The E-mode and B-mode patterns based on Eqs. (\ref{amp}), (\ref{angE}) and (
\ref{angB}) above are graphically represented in Figure \ref{fig4}; given
the polarization-ellipse angle $\psi _{\mathrm{o}}$, which determines the
orientation of the source frame $\left( x^{\prime },y^{\prime },z^{\prime
}\right) $ with respect to the observer's frame $\left( x,y,z\right) $ (see
Figure \ref{fig2} for comparison of the two frames). The patterns may be
characterized as follows.

\begin{enumerate}
\item[($\mathrm{i}$)] Figures \ref{fig:subfig1} (E-modes) and \ref%
{fig:subfig2} (B-modes) with $\psi _{\mathrm{o}}=0$:

The source frame $\left( x^{\prime },y^{\prime },z^{\prime }\right) $
coincides with the observer's frame $\left( x,y,z\right) $, and therefore
the rescattered EM radiation is received by the observer with its full
intensity and the maximum polarization effect. The E-modes and B-modes
exhibit regular gradient-like and curl-like patterns, respectively.

\item[($\mathrm{ii}$)] Figures \ref{fig:subfig3} (E-modes) and \ref%
{fig:subfig4} (B-modes) with $\psi _{\mathrm{o}}=\pi /4$:

The source frame $\left( x^{\prime },y^{\prime },z^{\prime }\right) $ is
rotated by $\pi /4$ from the observer's frame $\left( x,y,z\right) $ with
respect to the $z$-axis. The E-modes and B-modes exhibit the same patterns
as in case ($\mathrm{i}$) except that the both patterns are now shifted
along $\phi $ by $\psi _{\mathrm{o}}=\pi /4$; therefore physically
equivalent to case ($\mathrm{i}$).
\end{enumerate}

%TCIMACRO{%
%\TeXButton{TeX field}{\begin{figure}[tbp]
%\centering
%\subfloat[Subfigure 1 list of figures text][E-modes for $\psi _{\mathrm{o}}=0$]{
%\includegraphics[angle=270,width=0.46\textwidth]{E2-map-001}
%\label{fig:subfig1}} 
%\qquad 
%\subfloat[Subfigure 2 list of figures text][B-modes for $\psi _{\mathrm{o}}=0$]{
%\includegraphics[angle=270,width=0.46\textwidth]{B2-map-001}
%\label{fig:subfig2}} \ \ \ \ \ \ 
%\smallskip
%\subfloat[Subfigure 3 list of figures text][E-modes for $\psi _{\mathrm{o}}=\pi /4$]{
%\includegraphics[angle=270,width=0.46\textwidth]{E2-map-002}
%\label{fig:subfig3}} 
%\qquad 
%\subfloat[Subfigure 4 list of figures text][B-modes for $\psi _{\mathrm{o}}=\pi /4$]{
%\includegraphics[angle=270,width=0.46\textwidth]{B2-map-002}
%\label{fig:subfig4}} \ \ \ \ \ \ 
%\smallskip
%\caption{The E-mode and B-mode patterns of polarization given the polarization-ellipse angle $\psi _{\mathrm{o}}\
%$.}
%\label{fig4}
%\end{figure}}}%
%BeginExpansion
%EndExpansion

\subsection{Plane wave modulation of polarization patterns \label{plane}}

The E-mode and B-mode patterns of polarization in the previous Subsection
represent the local signature from scattering over the sphere. However, in
real-world observations the polarization patterns on the sky are not simply
the local signature from scattering but are modulated by plane wave
fluctuations on the last scattering surface. The plane wave modulation
changes the amplitude, sign and angular structure of the polarization but
not its nature: that is, it does not mix $Q$ and $U$ \cite{Hu}.

The quantities given by Eqs. (\ref{QpmU2}) or (\ref{QpmU3}) have a direct
association with physical observables, namely, Stokes parameters, and from
them we can illustrate the polarization patterns as was done at the end of
the previous Subsection. However, as can be seen from Eq. (\ref{QpmU3}),
these are spin $\pm 2$ quantities, and it would be desirable to derive
rotationally invariant spin $0$ quantities out of them \cite%
{zaldarriaga1997, kamionkowski1997, Kim}. Now, we define 
\begin{equation}
\tilde{Q}\pm \mathrm{i}\tilde{U}\equiv \left( \bar{Q}\pm \mathrm{i}\bar{U}%
\right) \exp \left( \mathrm{i}\mathbf{k}\cdot \mathbf{r}\right) =\left( \bar{%
Q}\pm \mathrm{i}\bar{U}\right) \exp \left( \mathrm{i}\rho \cos \theta
\right) ;~\rho \equiv kr,  \label{Qt}
\end{equation}%
where $\mathbf{k}=k\mathbf{e}_{z}$ and $\mathbf{r}=r\mathbf{e}_{r}$, with $k$
denoting the wave number and $r$ the comoving distance to the last
scattering surface, and $\exp \left( \mathrm{i}\mathbf{k}\cdot \mathbf{r}%
\right) =\exp \left( \mathrm{i}kr\cos \theta \right) =\exp \left( \mathrm{i}%
\rho \cos \theta \right) $ represents the plane wave projected into the
spherical sky \cite{Hu2}.\ From this we can construct the following scalar
quantities \cite{zaldarriaga1997, kamionkowski1997, Kim}: 
\begin{eqnarray}
\tilde{E} &\equiv &-\frac{1}{2}\left[ \bar{\eth }^{2}\left( \tilde{Q}+%
\mathrm{i}\tilde{U}\right) +\eth ^{2}\left( \tilde{Q}-\mathrm{i}\tilde{U}%
\right) \right] ,  \label{Et} \\
\tilde{B} &\equiv &\frac{\mathrm{i}}{2}\left[ \bar{\eth }^{2}\left( \tilde{Q}%
+\mathrm{i}\tilde{U}\right) -\eth ^{2}\left( \tilde{Q}-\mathrm{i}\tilde{U}%
\right) \right] ,  \label{Bt}
\end{eqnarray}%
where the spin raising and lowering operators are defined as 
\begin{eqnarray}
\eth \ _{s}f\left( \theta ,\phi \right) &=&-\left( \sin \theta \right)
^{s}\left( \frac{\partial }{\partial \theta }+\mathrm{i}\csc \theta \frac{%
\partial }{\partial \phi }\right) \left( \sin \theta \right) ^{-s}\
_{s}f\left( \theta ,\phi \right) ,  \label{eth} \\
\bar{\eth }\ _{s}f\left( \theta ,\phi \right) &=&-\left( \sin \theta \right)
^{-s}\left( \frac{\partial }{\partial \theta }-\mathrm{i}\csc \theta \frac{%
\partial }{\partial \phi }\right) \left( \sin \theta \right) ^{s}\
_{s}f\left( \theta ,\phi \right) ,  \label{ethb}
\end{eqnarray}%
for an arbitrary spin $s$ quantity $_{s}f\left( \theta ,\phi \right) $.
Substituting Eq. (\ref{QpmU2}) into Eq. (\ref{Qt}), and subsequently Eq. (%
\ref{Qt}) into Eqs. (\ref{Et}) and (\ref{Bt}), we obtain%
\begin{eqnarray}
\tilde{E} &=&\frac{3}{4}Q_{\mathrm{o}\,}\sin ^{2}\theta \left[ \rho
^{2}\left( 1+\cos ^{2}\theta \right) -12-8\mathrm{i}\rho \cos \theta \right]
\cos \left( 2\left( \phi -\psi _{\mathrm{o}}\right) \right) \exp \left( 
\mathrm{i}\rho \cos \theta \right) ,  \label{Et1} \\
\tilde{B} &=&\frac{3}{2}Q_{\mathrm{o}\,}\sin ^{2}\theta \left( -\rho
^{2}\cos \theta +4\mathrm{i}\rho \right) \sin \left( 2\left( \phi -\psi _{%
\mathrm{o}}\right) \right) \exp \left( \mathrm{i}\rho \cos \theta \right) .
\label{Bt1}
\end{eqnarray}%
Here we note that the spin $0$ quantity $\exp \left( \mathrm{i}\rho \cos
\theta \right) $ is a common phase factor for the complex quantities $\tilde{%
E}$ and $\tilde{B}$. With this factor suppressed, the relative phase between
the two complex quantities still remains the same. Then we may define 
\begin{eqnarray}
\bar{E} &\equiv &\tilde{E}\exp \left( -\mathrm{i}\rho \cos \theta \right) ,
\label{Eb} \\
\bar{B} &\equiv &\tilde{B}\exp \left( -\mathrm{i}\rho \cos \theta \right) ,
\label{Bb}
\end{eqnarray}%
such that their combinations are%
\begin{eqnarray}
\bar{E}+\mathrm{i}\bar{B} &=&\left( \tilde{E}+\mathrm{i}\tilde{B}\right)
\exp \left( -\mathrm{i}\rho \cos \theta \right) =-\bar{\eth }^{2}\left( 
\tilde{Q}+\mathrm{i}\tilde{U}\right) \exp \left( -\mathrm{i}\rho \cos \theta
\right) ,  \label{EpB} \\
\bar{E}-\mathrm{i}\bar{B} &=&\left( \tilde{E}-\mathrm{i}\tilde{B}\right)
\exp \left( -\mathrm{i}\rho \cos \theta \right) =-\eth ^{2}\left( \tilde{Q}-%
\mathrm{i}\tilde{U}\right) \exp \left( -\mathrm{i}\rho \cos \theta \right) ,
\label{EmB}
\end{eqnarray}%
by means of Eqs. (\ref{Et}) and (\ref{Bt}). By Eqs. (\ref{Et1}) and (\ref%
{Bt1}) these read 
\begin{eqnarray}
\bar{E}\pm \mathrm{i}\bar{B} &=&\frac{3}{4}Q_{\mathrm{o}\,}\sin ^{2}\theta
\left\{ \left[ \rho ^{2}\left( 1+\cos ^{2}\theta \right) -12\right] \cos
\left( 2\left( \phi -\psi _{\mathrm{o}}\right) \right) \mp \ 8\rho \sin
\left( 2\left( \phi -\psi _{\mathrm{o}}\right) \right) \right.  \notag \\
&&\left. +\ \mathrm{i}\cos \theta \left[ -8\rho \cos \left( 2\left( \phi
-\psi _{\mathrm{o}}\right) \right) \mp 2\rho ^{2}\sin \left( 2\left( \phi
-\psi _{\mathrm{o}}\right) \right) \right] \right\} .  \label{EpmB}
\end{eqnarray}%
Here we note that $\bar{E}-\mathrm{i}\bar{B}$ can be obtained by changing $\phi -\psi _{\mathrm{o}}\rightarrow -\left( \phi -\psi _{\mathrm{o}}\right) $
in $\bar{E}+\mathrm{i}\bar{B}$. This means that $\bar{E}-\mathrm{i}\bar{B}$
is the reflection of$\ \bar{E}+\mathrm{i}\bar{B}$ about $\phi =\psi _{\mathrm{o}}$. Therefore, $\bar{E}+\mathrm{i}\bar{B}$ and $\bar{E}-\mathrm{i} \bar{B}$ lead to physically equivalent representations: they produce
patterns which are simply mirror images of each other \footnote{The same argument applies to $\bar{Q}+\mathrm{i}\bar{U}$ and $\bar{Q}-\mathrm{i}\bar{U}$ due to Eq. (\ref{QpmU2}).}.

We choose $\bar{E}+\mathrm{i}\bar{B}$ for our representation of polarization
patterns. Similarly as in the previous Subsection, one can illustrate the
E-mode and\ B-mode polarization patterns by means of Eq. (\ref{EpmB}). We
represent E-modes by the polarization amplitude and angle clockwise from the
north pole (at $\theta =0$), and B-modes by rotating the E-modes by 45$^{\circ }$:
\begin{equation}
\text{\textrm{Amplitude}}=\sqrt{\left[ \Re \left( \bar{E}+\mathrm{i}\bar{B}%
\right) \right] ^{2}+\left[ \Im \left( \bar{E}+\mathrm{i}\bar{B}\right) %
\right] ^{2}},  \label{amp1}
\end{equation}%
\begin{eqnarray}
\mathrm{Angle}\left( \text{\textrm{E-modes}}\right) &=&\frac{1}{2}\arctan
\left( \frac{\Im \left( \bar{E}+\mathrm{i}\bar{B}\right) }{\Re \left( \bar{E}%
+\mathrm{i}\bar{B}\right) }\right) ,  \label{angE1} \\
\mathrm{Angle}\left( \text{\textrm{B-modes}}\right) &=&\mathrm{Angle}\left( 
\text{\textrm{E-modes}}\right) +\frac{\pi }{4},  \label{angB1}
\end{eqnarray}%
where 
\begin{eqnarray}
\Re \left( \bar{E}+\mathrm{i}\bar{B}\right) &=&\frac{3}{4}Q_{\mathrm{o}%
\,}\sin ^{2}\theta \left\{ \left[ \rho ^{2}\left( 1+\cos ^{2}\theta \right)
-12\right] \cos \left( 2\left( \phi -\psi _{\mathrm{o}}\right) \right) -\
8\rho \sin \left( 2\left( \phi -\psi _{\mathrm{o}}\right) \right) \right\} ,
\label{re} \\
\Im \left( \bar{E}+\mathrm{i}\bar{B}\right) &=&-\frac{3}{2}Q_{\mathrm{o}%
\,}\sin ^{2}\theta \cos \theta \left[ 4\rho \cos \left( 2\left( \phi -\psi _{%
\mathrm{o}}\right) \right) +\rho ^{2}\sin \left( 2\left( \phi -\psi _{%
\mathrm{o}}\right) \right) \right] ,  \label{im}
\end{eqnarray}%
with $Q_{\mathrm{o}\,}$ given by Eq. (\ref{Q04}).

In particular, for $\rho \gg 1$, which corresponds to small angular scales,
the representations of E-modes and\ B-modes tend to%
\begin{equation}
\text{\textrm{Amplitude}}=\frac{3}{4}\left\vert Q_{\mathrm{o}\,}\right\vert
\rho ^{2}\sin ^{2}\theta \left\{ \left[ \left( 1+\cos ^{2}\theta \right)
\cos \left( 2\left( \phi -\psi _{\mathrm{o}}\right) \right) \right] ^{2}+%
\left[ -2\cos \theta \sin \left( 2\left( \phi -\psi _{\mathrm{o}}\right)
\right) \right] ^{2}\right\} ^{1/2},  \label{amp2}
\end{equation}%
\begin{eqnarray}
\mathrm{Angle}\left( \text{\textrm{E-modes}}\right) &=&\frac{1}{2}\arctan
\left( \frac{-2\cos \theta \tan \left( 2\left( \phi -\psi _{\mathrm{o}%
}\right) \right) }{\left( 1+\cos ^{2}\theta \right) }\right) ,  \label{angE2}
\\
\mathrm{Angle}\left( \text{\textrm{B-modes}}\right) &=&\mathrm{Angle}\left( 
\text{\textrm{E-modes}}\right) +\frac{\pi }{4}.  \label{angB2}
\end{eqnarray}%
We note that these scale to the spin $2$ representations given by Eqs. (\ref%
{amp}), (\ref{angE}) and (\ref{angB}), apart from the factor $\rho ^{2}\sin
^{2}\theta $ in the amplitude \cite{Chiang}.

\begin{center}
\textbf{Graphic representation 2}
\end{center}

%TCIMACRO{%
%\TeXButton{TeX field}{\begin{figure}[tbp]
%\centering
%\subfloat[Subfigure 1 list of figures text][E-modes for $\rho=5\ $and $\psi _{\mathrm{o}}=0$]{
%\includegraphics[angle=270,width=0.46\textwidth]{E05-map-002}
%\label{fig:subfig9}} 
%\qquad 
%\subfloat[Subfigure 2 list of figures text][B-modes for $\rho=5\ $and $\psi _{\mathrm{o}}=0$]{
%\includegraphics[angle=270,width=0.46\textwidth]{B05-map-002}
%\label{fig:subfig10}} \ \ \ \ \ \ 
%\smallskip
%\subfloat[Subfigure 3 list of figures text][E-modes for $\rho=5\ $and $\psi _{\mathrm{o}}=\pi /4$]{
%\includegraphics[angle=270,width=0.46\textwidth]{E05-map-006}
%\label{fig:subfig11}} 
%\qquad 
%\subfloat[Subfigure 4 list of figures text][B-modes for $\rho=5\ $and $\psi _{\mathrm{o}}=\pi /4$]{
%\includegraphics[angle=270,width=0.46\textwidth]{B05-map-006}
%\label{fig:subfig12}} \ \ \ \ \ \ 
%\smallskip
%\caption{The E-mode and B-mode patterns of polarization for $\rho=5\ $and the polarization-ellipse angle $\psi _{\mathrm{o}}\
%$.}
%\label{fig5}
%\end{figure}}}%
%BeginExpansion
\begin{figure}[tbp]
\centering
\subfloat[Subfigure 1 list of figures text][E-modes for $\rho=5\ $and $\psi _{\mathrm{o}}=0$]{
\includegraphics[angle=270,width=0.46\textwidth]{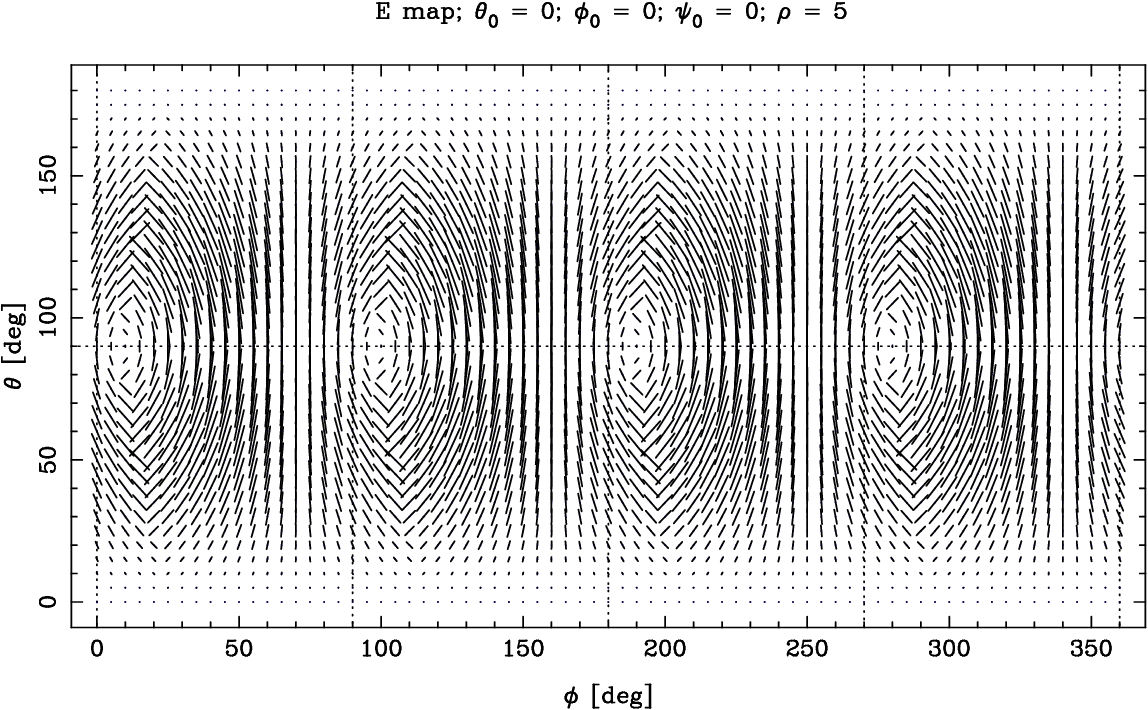}
\label{fig:subfig9}} 
\qquad 
\subfloat[Subfigure 2 list of figures text][B-modes for $\rho=5\ $and $\psi _{\mathrm{o}}=0$]{
\includegraphics[angle=270,width=0.46\textwidth]{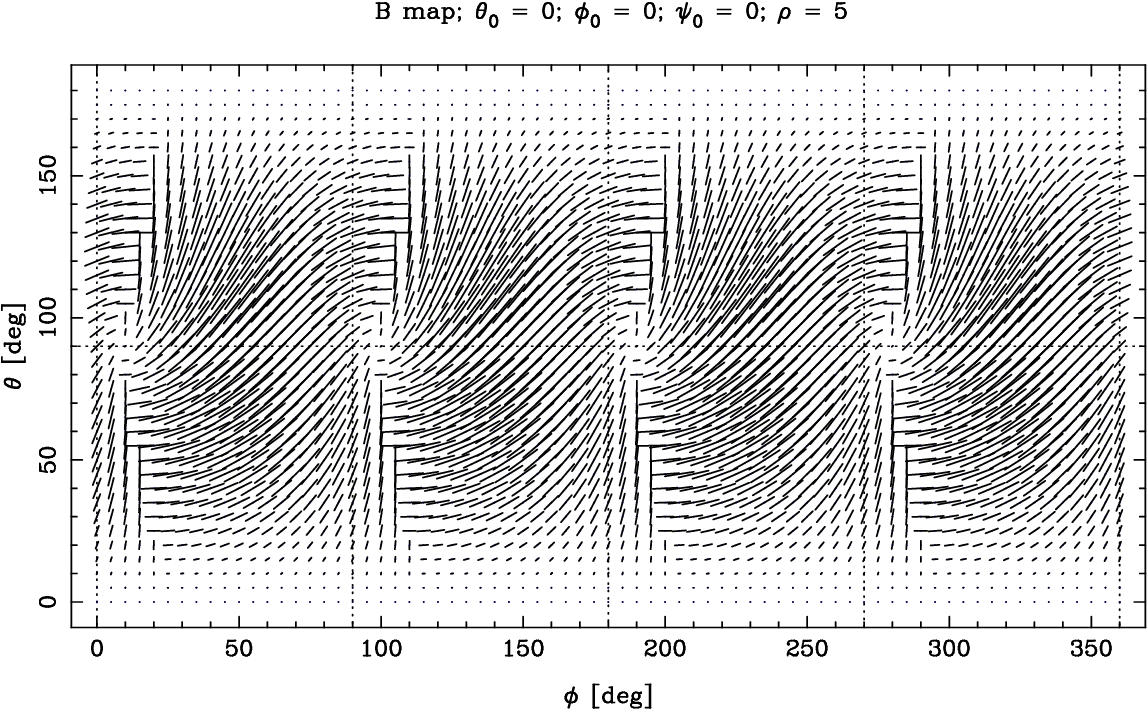}
\label{fig:subfig10}} \ \ \ \ \ \ 
\smallskip
\subfloat[Subfigure 3 list of figures text][E-modes for $\rho=5\ $and $\psi _{\mathrm{o}}=\pi /4$]{
\includegraphics[angle=270,width=0.46\textwidth]{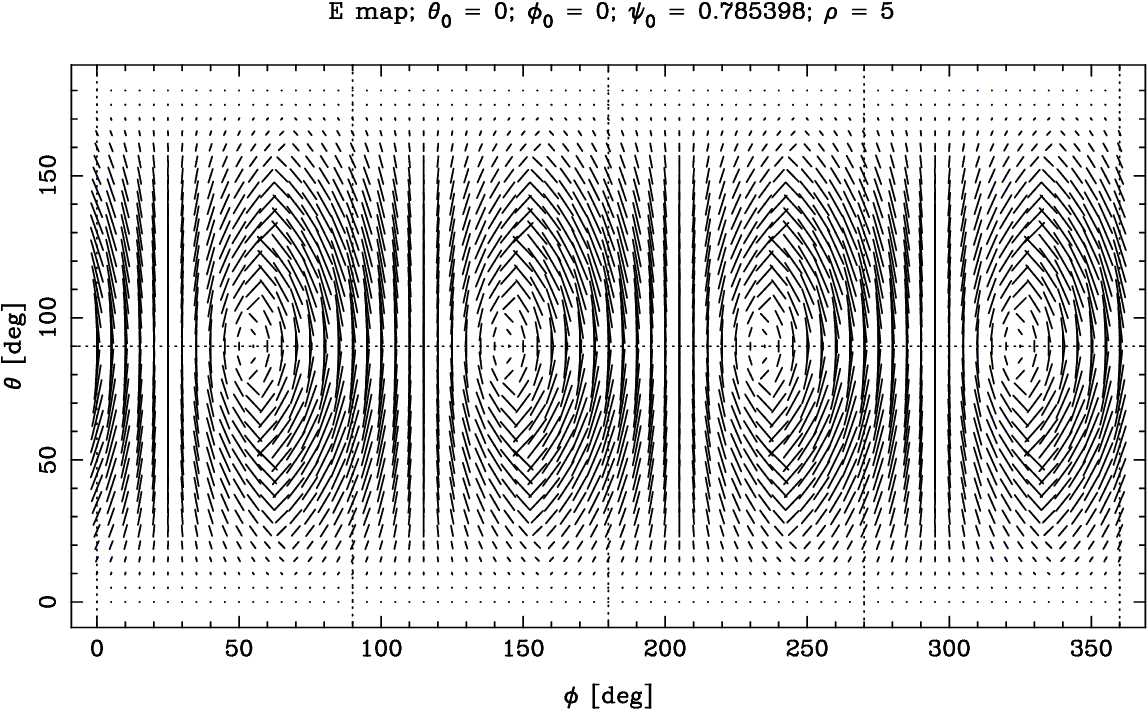}
\label{fig:subfig11}} 
\qquad 
\subfloat[Subfigure 4 list of figures text][B-modes for $\rho=5\ $and $\psi _{\mathrm{o}}=\pi /4$]{
\includegraphics[angle=270,width=0.46\textwidth]{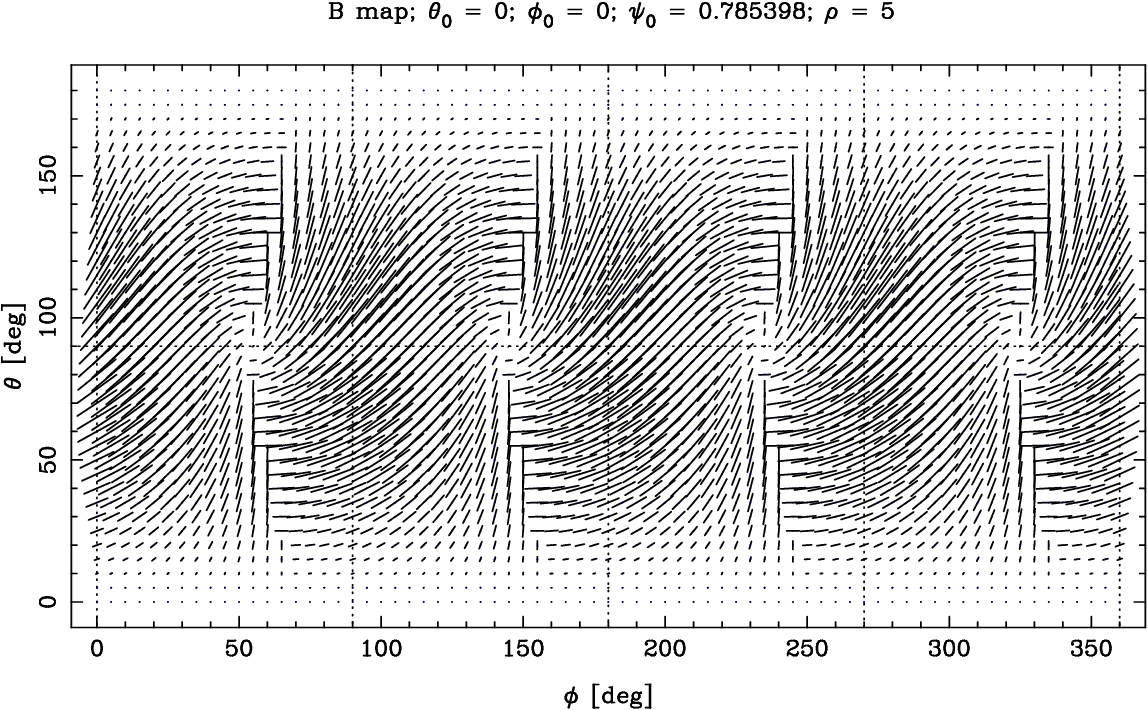}
\label{fig:subfig12}} \ \ \ \ \ \ 
\smallskip
\caption{The E-mode and B-mode patterns of polarization for $\rho=5\ $and the polarization-ellipse angle $\psi _{\mathrm{o}}\
$.}
\label{fig5}
\end{figure}%
%EndExpansion

%TCIMACRO{%
%\TeXButton{TeX field}{\begin{figure}[tbp]
%\centering
%\subfloat[Subfigure 1 list of figures text][E-modes for $\rho=25\ $and $\psi _{\mathrm{o}}=0$]{
%\includegraphics[angle=270,width=0.46\textwidth]{E025-map-003}
%\label{fig:subfig17}} 
%\qquad 
%\subfloat[Subfigure 2 list of figures text][B-modes for $\rho=25\ $and $\psi _{\mathrm{o}}=0$]{
%\includegraphics[angle=270,width=0.46\textwidth]{B025-map-003}
%\label{fig:subfig18}} \ \ \ \ \ \ 
%\smallskip
%\subfloat[Subfigure 3 list of figures text][E-modes for $\rho=25\ $and $\psi _{\mathrm{o}}=\pi /4$]{
%\includegraphics[angle=270,width=0.46\textwidth]{E025-map-007}
%\label{fig:subfig19}} 
%\qquad 
%\subfloat[Subfigure 4 list of figures text][B-modes for $\rho=25\ $and $\psi _{\mathrm{o}}=\pi /4$]{
%\includegraphics[angle=270,width=0.46\textwidth]{B025-map-007}
%\label{fig:subfig20}} \ \ \ \ \ \ 
%\smallskip
%\caption{The E-mode and B-mode patterns of polarization for $\rho=25\ $and the polarization-ellipse angle $\psi _{\mathrm{o}}\
%$.}
%\label{fig6}
%\end{figure}}}%
%BeginExpansion
\begin{figure}[tbp]
\centering
\subfloat[Subfigure 1 list of figures text][E-modes for $\rho=25\ $and $\psi _{\mathrm{o}}=0$]{
\includegraphics[angle=270,width=0.46\textwidth]{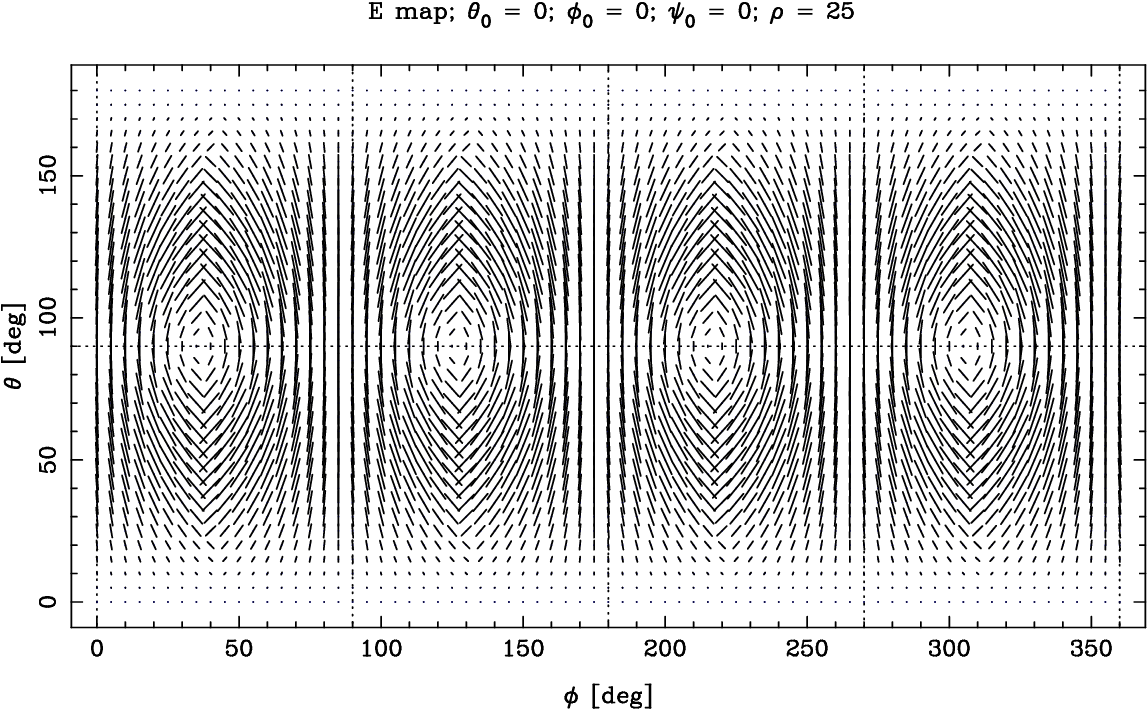}
\label{fig:subfig17}} 
\qquad 
\subfloat[Subfigure 2 list of figures text][B-modes for $\rho=25\ $and $\psi _{\mathrm{o}}=0$]{
\includegraphics[angle=270,width=0.46\textwidth]{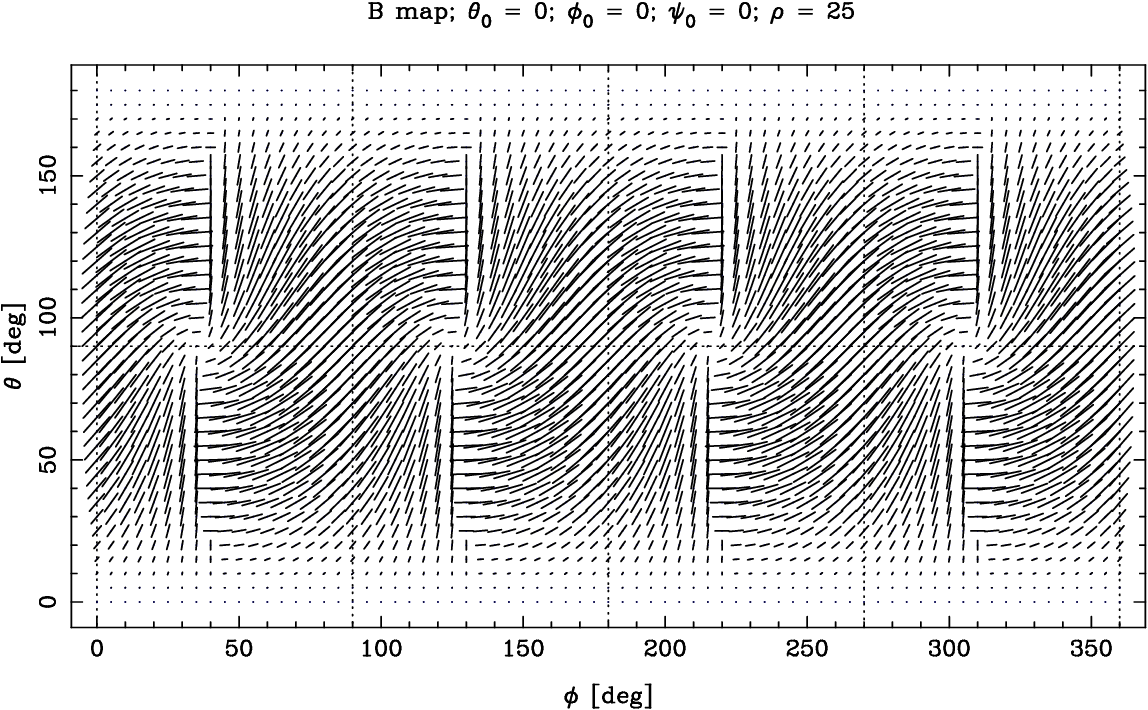}
\label{fig:subfig18}} \ \ \ \ \ \ 
\smallskip
\subfloat[Subfigure 3 list of figures text][E-modes for $\rho=25\ $and $\psi _{\mathrm{o}}=\pi /4$]{
\includegraphics[angle=270,width=0.46\textwidth]{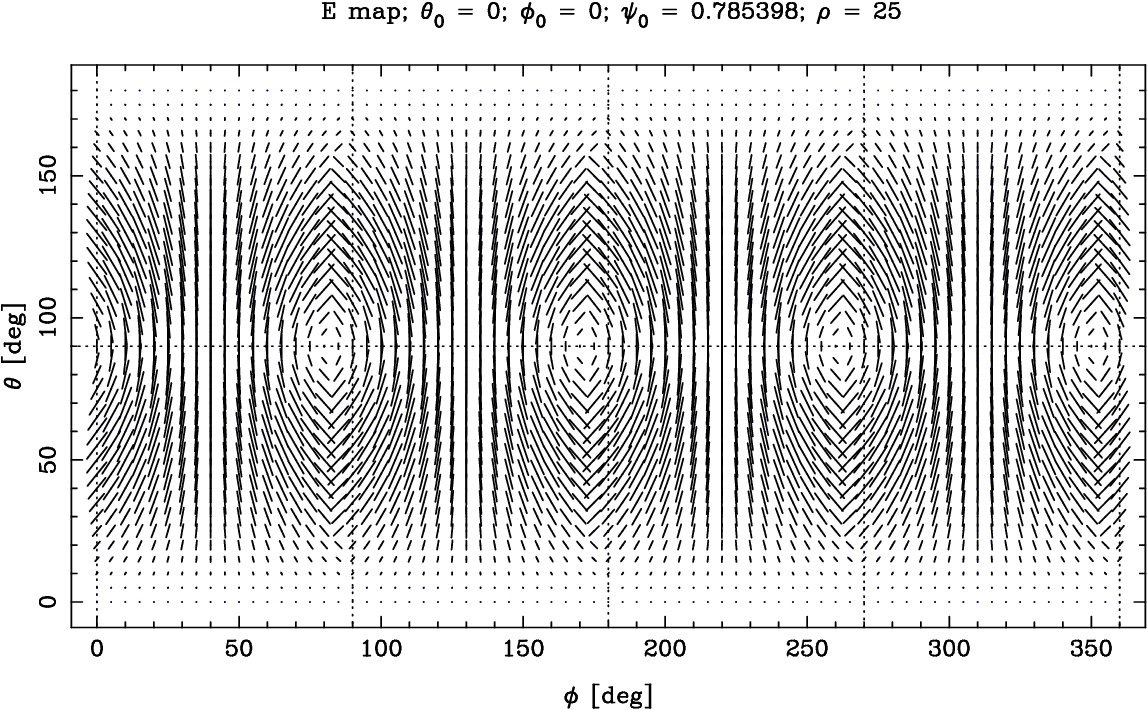}
\label{fig:subfig19}} 
\qquad 
\subfloat[Subfigure 4 list of figures text][B-modes for $\rho=25\ $and $\psi _{\mathrm{o}}=\pi /4$]{
\includegraphics[angle=270,width=0.46\textwidth]{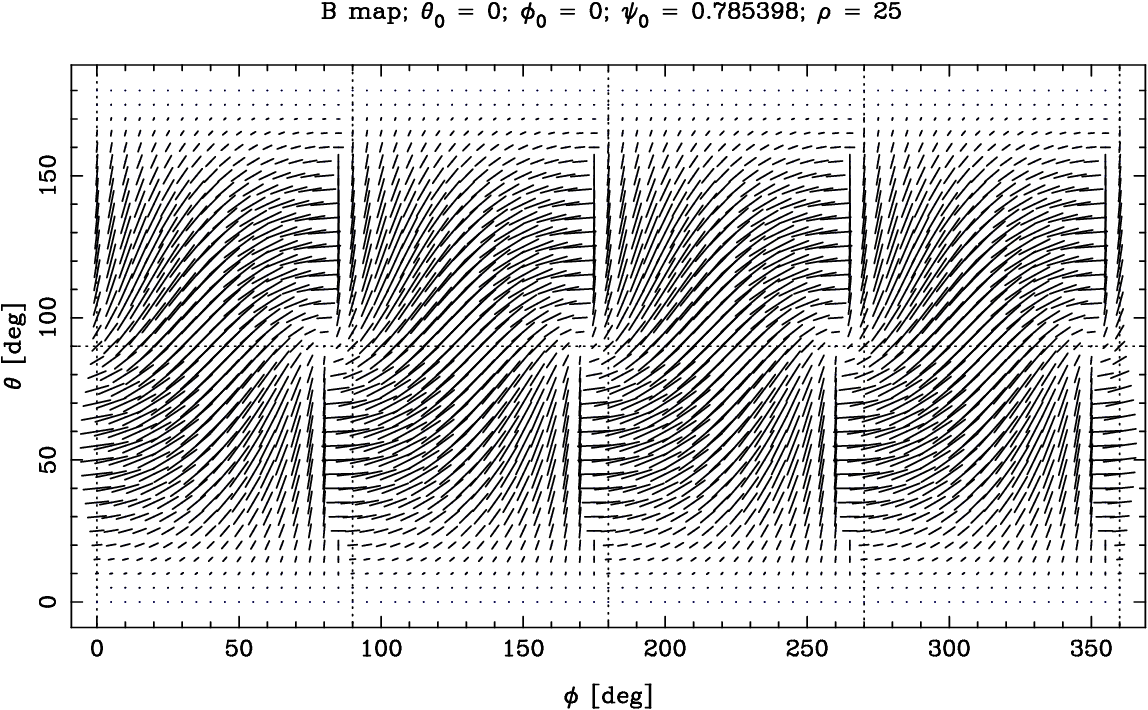}
\label{fig:subfig20}} \ \ \ \ \ \ 
\smallskip
\caption{The E-mode and B-mode patterns of polarization for $\rho=25\ $and the polarization-ellipse angle $\psi _{\mathrm{o}}\
$.}
\label{fig6}
\end{figure}%
%EndExpansion

%TCIMACRO{%
%\TeXButton{TeX field}{\begin{figure}[tbp]
%\centering
%\subfloat[Subfigure 1 list of figures text][E-modes for $\rho=125\ $and $\psi _{\mathrm{o}}=0$]{
%\includegraphics[angle=270,width=0.46\textwidth]{E0125-map-004}
%\label{fig:subfig25}} 
%\qquad 
%\subfloat[Subfigure 2 list of figures text][B-modes for $\rho=125\ $and $\psi _{\mathrm{o}}=0$]{
%\includegraphics[angle=270,width=0.46\textwidth]{B0125-map-004}
%\label{fig:subfig26}} \ \ \ \ \ \ 
%\smallskip
%\subfloat[Subfigure 3 list of figures text][E-modes for $\rho=125\ $and $\psi _{\mathrm{o}}=\pi /4$]{
%\includegraphics[angle=270,width=0.46\textwidth]{E0125-map-008}
%\label{fig:subfig27}} 
%\qquad 
%\subfloat[Subfigure 4 list of figures text][B-modes for $\rho=125\ $and $\psi _{\mathrm{o}}=\pi /4$]{
%\includegraphics[angle=270,width=0.46\textwidth]{B0125-map-008}
%\label{fig:subfig28}} \ \ \ \ \ \ 
%\smallskip
%\caption{The E-mode and B-mode patterns of polarization for $\rho=125\ $and the polarization-ellipse angle $\psi _{\mathrm{o}}\
%$.}
%\label{fig7}
%\end{figure}}}%
%BeginExpansion
\begin{figure}[tbp]
\centering
\subfloat[Subfigure 1 list of figures text][E-modes for $\rho=125\ $and $\psi _{\mathrm{o}}=0$]{
\includegraphics[angle=270,width=0.46\textwidth]{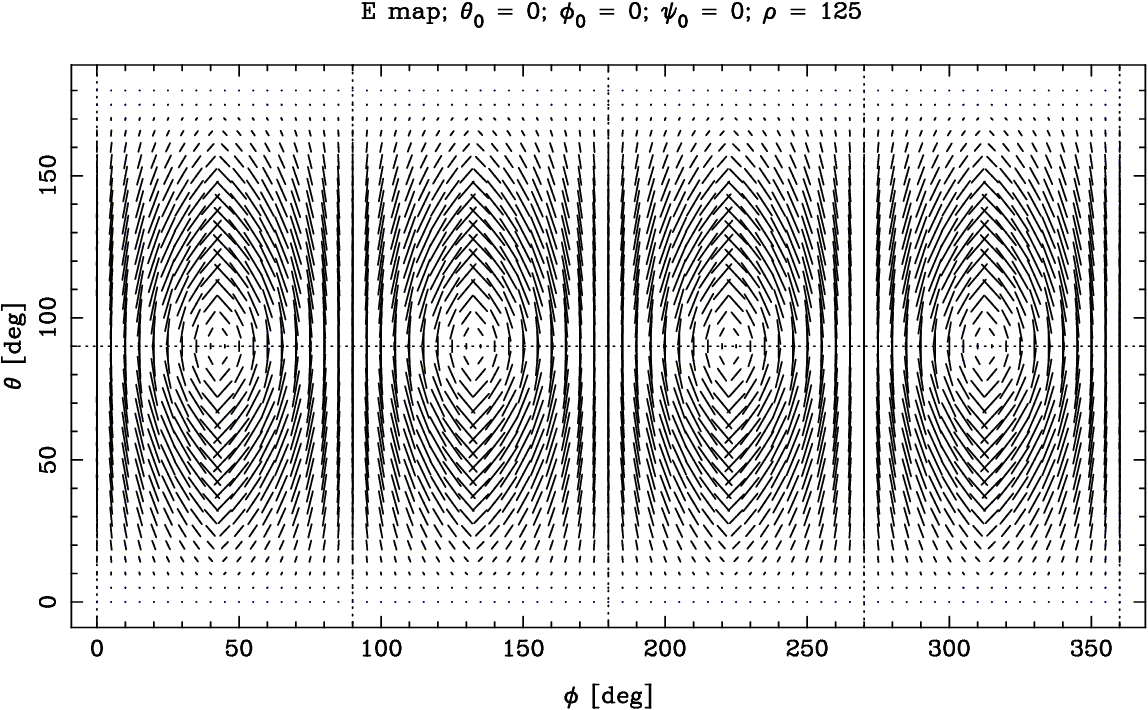}
\label{fig:subfig25}} 
\qquad 
\subfloat[Subfigure 2 list of figures text][B-modes for $\rho=125\ $and $\psi _{\mathrm{o}}=0$]{
\includegraphics[angle=270,width=0.46\textwidth]{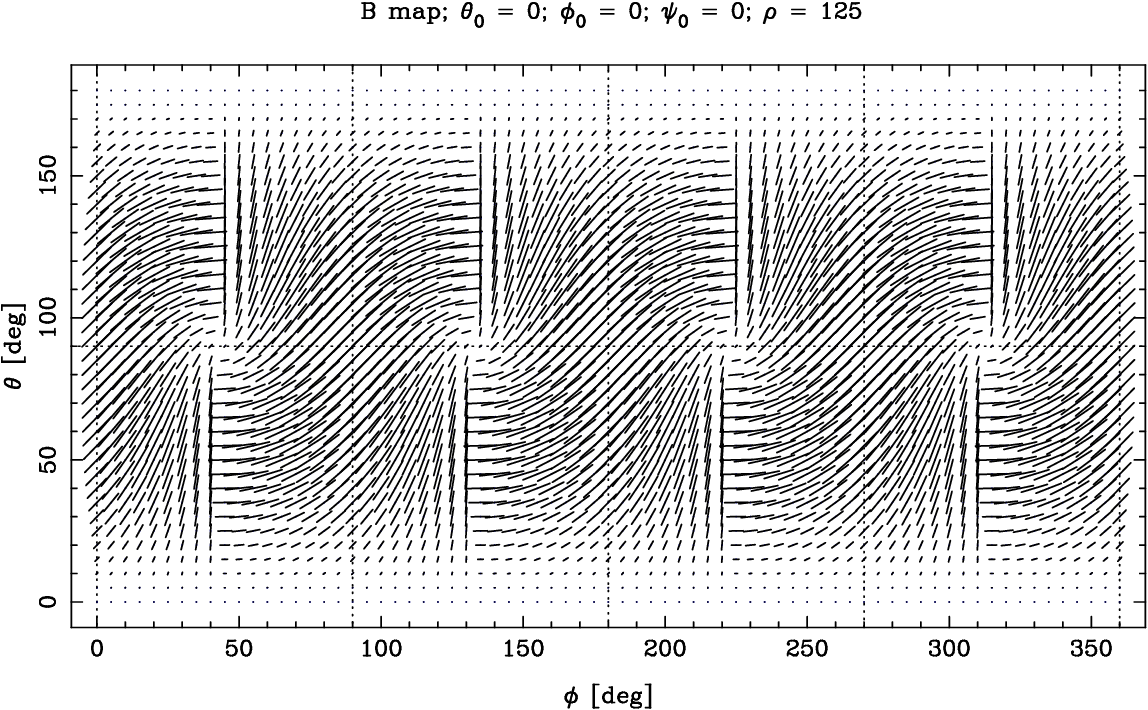}
\label{fig:subfig26}} \ \ \ \ \ \ 
\smallskip
\subfloat[Subfigure 3 list of figures text][E-modes for $\rho=125\ $and $\psi _{\mathrm{o}}=\pi /4$]{
\includegraphics[angle=270,width=0.46\textwidth]{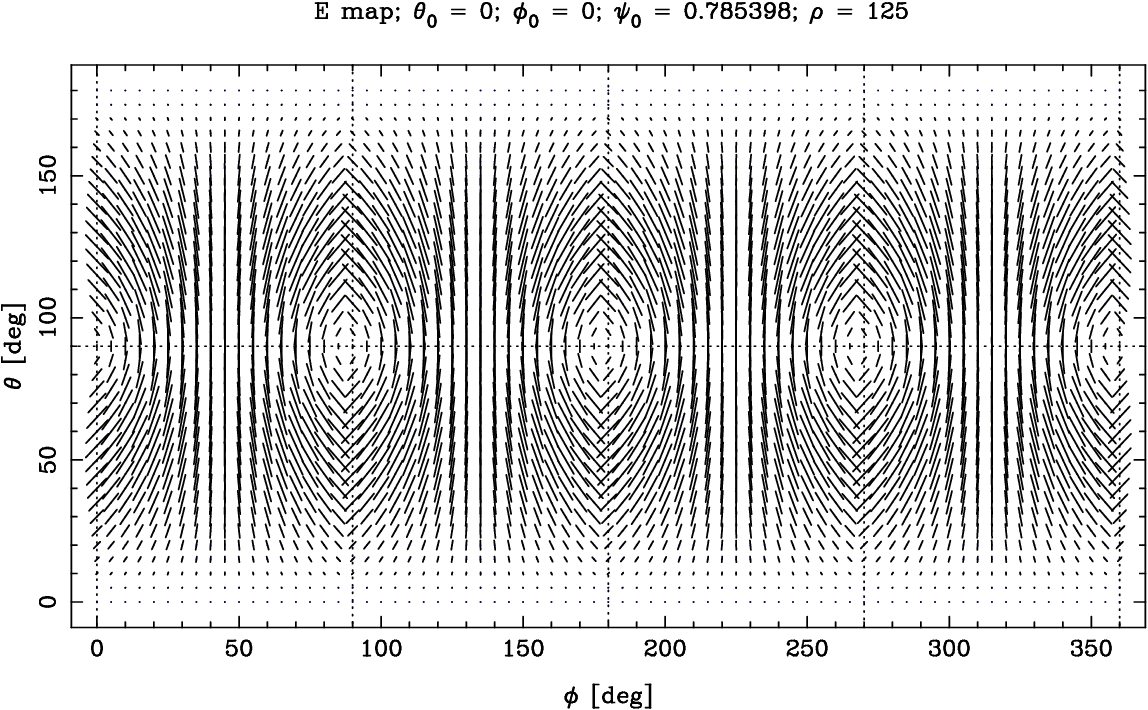}
\label{fig:subfig27}} 
\qquad 
\subfloat[Subfigure 4 list of figures text][B-modes for $\rho=125\ $and $\psi _{\mathrm{o}}=\pi /4$]{
\includegraphics[angle=270,width=0.46\textwidth]{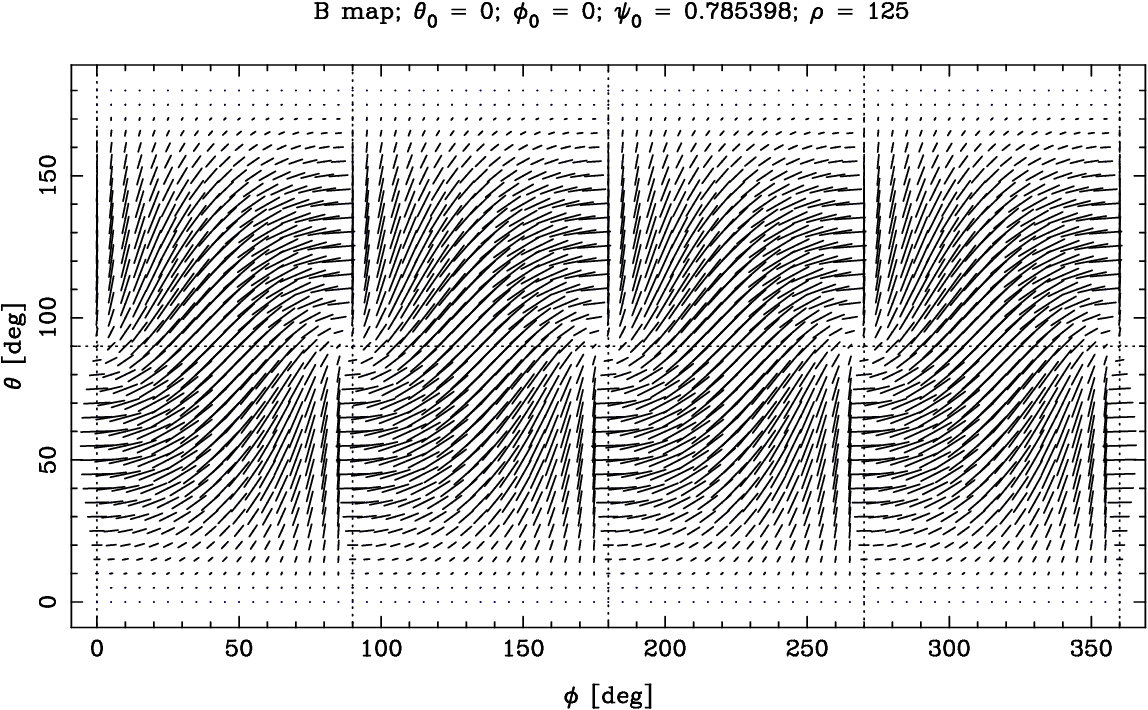}
\label{fig:subfig28}} \ \ \ \ \ \ 
\smallskip
\caption{The E-mode and B-mode patterns of polarization for $\rho=125\ $and the polarization-ellipse angle $\psi _{\mathrm{o}}\
$.}
\label{fig7}
\end{figure}%
%EndExpansion

The E-mode and B-mode patterns based on Eqs. (\ref{amp1}), (\ref{angE1}) and
(\ref{angB1}) above are graphically represented in Figures \ref{fig5}, \ref%
{fig6} and \ref{fig7}; with the dimensionless parameter $\rho $, which
defines the plane wave projection into the spherical sky (i.e. $\exp \left( 
\mathrm{i}\mathbf{k}\cdot \mathbf{r}\right) =\exp \left( \mathrm{i}kr\cos
\theta \right) =\exp \left( \mathrm{i}\rho \cos \theta \right) $) and with
the polarization-ellipse angle $\psi _{\mathrm{o}}$, which determines the
orientation of the source frame $\left( x^{\prime },y^{\prime },z^{\prime
}\right) $ with respect to the observer's frame $\left( x,y,z\right) $ (see
Figure \ref{fig2} for comparison of the two frames). Given different values
of $\rho $ and $\psi _{\mathrm{o}}$, the patterns may be characterized as
follows.

\begin{enumerate}
\item[\textit{1.}] \textit{In Figure \ref{fig5}}

\item[($\mathrm{i}^{\prime }$)] Figures \ref{fig:subfig9} (E-modes) and \ref%
{fig:subfig10} (B-modes) with $\rho =5$, $\psi _{\mathrm{o}}=0$:

The source frame $\left( x^{\prime },y^{\prime },z^{\prime }\right) $
coincides with the observer's frame $\left( x,y,z\right) $, and therefore
the rescattered EM radiation is received by the observer with its full
intensity and the maximum polarization effect. The E-modes and B-modes
exhibit gradient-like and curl-like patterns but with broken symmetric and
anti-symmetric images in each quadrant for $\phi $, respectively. This is
caused by the plane wave modulation at \textit{long} wavelengths or at 
\textit{small} comoving distances to the last scattering surface; due to
Eqs. (\ref{angE1}) and (\ref{angB1}) with $\rho =2\pi r/\lambda $.

\item[(i$\mathrm{i}^{\prime }$)] Figures \ref{fig:subfig11} (E-modes) and %
\ref{fig:subfig12} (B-modes) with $\rho =5$, $\psi _{\mathrm{o}}=\pi /4$:

The source frame $\left( x^{\prime },y^{\prime },z^{\prime }\right) $ is
rotated by $\pi /4$ from the observer's frame $\left( x,y,z\right) $ with
respect to the $z$-axis. The E-modes and B-modes exhibit the same patterns
as in case ($\mathrm{i}^{\prime }$) except that the both patterns are now
shifted along $\phi $ by $\psi _{\mathrm{o}}=$ $\pi /4$; therefore
physically equivalent to case ($\mathrm{i}^{\prime }$).

\item[\textit{2.}] \textit{In Figure \ref{fig6}}

\item[($\mathrm{i}^{\prime \prime }$)] Figures \ref{fig:subfig17} (E-modes)
and \ref{fig:subfig18} (B-modes) with $\rho =25$, $\psi _{\mathrm{o}}=0$:

The source frame $\left( x^{\prime },y^{\prime },z^{\prime }\right) $
coincides with the observer's frame $\left( x,y,z\right) $, and therefore
the rescattered EM radiation is received by the observer with its full
intensity and the maximum polarization effect. The E-modes and B-modes
exhibit gradient-like and curl-like patterns but with broken symmetric and
anti-symmetric images in each quadrant for $\phi $, respectively; however,
less broken than in case ($\mathrm{i}^{\prime }$). This is caused by the
plane wave modulation at \textit{medium} wavelengths or at \textit{medium}
comoving distances to the last scattering surface; due to Eqs. (\ref{angE1})
and (\ref{angB1}) with $\rho =2\pi r/\lambda $.

\item[(i$\mathrm{i}^{\prime \prime }$)] Figures \ref{fig:subfig19} (E-modes)
and \ref{fig:subfig20} (B-modes) with $\rho =25$, $\psi _{\mathrm{o}}=\pi /4$%
:

The source frame $\left( x^{\prime },y^{\prime },z^{\prime }\right) $ is
rotated by $\pi /4$ from the observer's frame $\left( x,y,z\right) $ with
respect to the $z$-axis. The E-modes and B-modes exhibit the same patterns
as in case ($\mathrm{i}^{\prime \prime }$) except that the both patterns are
now shifted along $\phi $ by $\psi _{\mathrm{o}}=$ $\pi /4$; therefore
physically equivalent to case ($\mathrm{i}^{\prime \prime }$).

\item[\textit{3.}] \textit{In Figure \ref{fig7}}

\item[($\mathrm{i}^{\prime \prime \prime }$)] Figures \ref{fig:subfig25}
(E-modes) and \ref{fig:subfig26} (B-modes) with $\rho =125$, $\psi _{\mathrm{
o}}=0$:

The source frame $\left( x^{\prime },y^{\prime },z^{\prime }\right) $
coincides with the observer's frame $\left( x,y,z\right) $, and therefore
the rescattered EM radiation is received by the observer with its full
intensity and the maximum polarization effect. The E-modes and B-modes
exhibit \textit{regular} gradient-like and curl-like patterns with proper
symmetric and anti-symmetric images in each quadrant for $\phi $,
respectively. This is caused by the plane wave modulation at \textit{short}
wavelengths or at \textit{large} comoving distances to the last scattering
surface; due to Eqs. (\ref{angE1}) and (\ref{angB1}) with $\rho =2\pi
r/\lambda $.

\item[(i$\mathrm{i}^{\prime \prime \prime }$)] Figures \ref{fig:subfig27}
(E-modes) and \ref{fig:subfig28} (B-modes) with $\rho =125$, $\psi _{\mathrm{%
o}}=\pi /4$:

The source frame $\left( x^{\prime },y^{\prime },z^{\prime }\right) $ is
rotated by $\pi /4$ from the observer's frame $\left( x,y,z\right) $ with
respect to the $z$-axis. The E-modes and B-modes exhibit the same patterns
as in case ($\mathrm{i}^{\prime \prime \prime }$) except that the both
patterns are now shifted along $\phi $ by $\psi _{\mathrm{o}}=$ $\pi /4$;
therefore physically equivalent to case ($\mathrm{i}^{\prime \prime \prime }$%
).
\end{enumerate}

%\newline

Cases ($\mathrm{i}^{\prime }$) and ($\mathrm{i}^{\prime \prime }$) from
Figures \ref{fig5} and \ref{fig6} show that the broken symmetric and
anti-symmetric images in the E-mode and B-mode patterns are due to the plane
wave modulation at \textit{long} to \textit{medium} wavelengths or at 
\textit{small} to \textit{medium} comoving distances to the last scattering
surface, which corresponds to small to medium $\rho $. However, case ($%
\mathrm{i}^{\prime \prime \prime }$) from Figure \ref{fig7} shows that the
E-mode and B-mode patterns take proper symmetric and anti-symmetric images,
respectively, with the plane wave modulation at \textit{short} wavelengths
or at \textit{large} comoving distances to the last scattering surface,
which corresponds to large $\rho $: in fact, these patterns should be
equivalent to the ones to be produced via Eqs. (\ref{amp2}),\ (\ref{angE2})
and (\ref{angB2}) in the limit $\rho \gg 1$. While the plane wave modulation
affects the symmetric and anti-symmetric properties of the E-modes and
B-modes, depending on the size of $\rho $, it does not change the
periodicity of the patterns in $\phi $ (i.e. $\pi /2$): in both the E-mode
and B-mode patterns, four identical images are produced over $0\leq \phi
<2\pi $, one in each of the four quadrants.

\section{Discussion and Conclusions}

We present an analysis of an apparently as yet overlooked effect in the interaction between (primordial) GWs and individual electric
charges. We computed the Stokes parameters for the EM radiation rescattered
by a charge that is forced into oscillatory motion by GWs propagating
through the CMB plasma: as shown by Eqs. (\ref{I01}) -- (\ref{V01}) and (\ref{i0}) -- (\ref{v0}) in Section \ref{Stokes}, the Stokes parameters exhibit a
net linear polarization -- as expected. 
The GWs \emph{open polarization channels} for the rescattered
radiation by setting a charge in motion along particular directions. In
other words, the GWs provide for polarization in
the photon-electron scattering by agitating a charge along particular
directions.
Taking the $\mathcal{O}\left( h\right) $ terms into consideration,
the resulting contribution to our Stokes parameters will be as shown in Eqs.
(\ref{Is}) -- (\ref{Vs}), which is $\mathcal{O}\left( h^{2}\right) $. In our analysis,
due to the assumption of linearized gravity in Section II A, $\mathcal{O}
\left(h^{2}\right)$ would be considered as very small, and therefore the
terms of this order may be disregarded for the final expressions of the
Stokes parameters as in Eqs. (\ref{I01}) -- (\ref{V01}). However,
until we are able to estimate how large or small the strain $h$ is in the CMB
regime [unlike its estimates for present time observations as in Figure 3],
it is not clear whether the negligence of $\mathcal{O}\left(
h^{2}\right) $ will make any significant difference (or not) in our final
E-mode/B-mode maps. Our analysis is presently limited to modeling single-electron scattering; accordingly, we are not yet able to provide a reliable estimate of the amount of polarized flux to be expected. Evidently, reliable amplitude estimates will be crucial for attempts to observe the effect we describe in this study; current B-mode polarization studies have sensitivities of about $0.3\,\mu{\rm K}\sqrt{s}$ (in CMB temperature units, using BICEP2 as benchmark; \cite{bicep2014}).

If our
measurement were indeed sensitive enough to tell the values of the strain $h$ and the
frequency $\omega $ of the GWs in the CMB regime, then we should rather
express our Stokes parameters using Eqs. (\ref{Is}) -- (\ref{Vs}). This way, we would be
able to see the properties of the GWs directly from our E-mode/B-mode
polarization maps through the Stokes parameters. In particular, the second
parameter $Q$ would take different values, depending on the GW polarization
states, $+$ and $\times $, while the first parameter $I$ would be the same
regardless of the states. This implies that the linear polarization of our
CMB radiation would be quantified differently, depending on the GW
polarization states. That is, $Q_{+}=Q_{\mathrm{o}}+\mathcal{O}\left(
h^{2}\right) $ and $Q_{\times }=Q_{\mathrm{o}}-\mathcal{O}\left(
h^{2}\right) $ while $I_{+}=I_{\times }=I_{\mathrm{o}}+\mathcal{O}\left(
h^{2}\right) $, and therefore we find the ratios, $Q_{+}/I_{+}\neq Q_{\times
}/I_{\times }$. This means that the GW-induced polarization should show this
distinctive feature, which will distinguish itself from other degenerate
cases, where polarization could possibly have been induced from non-GW
sources.

In order to analyze the CMB
polarization on the sky, we have redefined the Stokes parameters on a sphere
such that their representations can be expressed as spin-weighted spherical
harmonics; namely, spin $\pm 2$ quantities. The representations yield the
two polarization patterns, \textquotedblleft electric\textquotedblright\
E-mode and \textquotedblleft magnetic\textquotedblright\ B-mode patterns 
\cite{Hu}. They are constructed by means of Eqs. (\ref{amp}), (\ref{angE})
and (\ref{angB}), and graphically represented in Figure \ref{fig4} (see 
\textbf{Graphic representations 1}) in Section \ref{sphere}. While these
patterns represent the local signature from scattering over the sphere, in
real-world observations, the polarization patterns on the sky are not simply
the local signature from scattering but are modulated by plane wave
fluctuations on the last scattering surface \cite{Hu}. Thus we have modified
the representations of the Stokes parameters such that they contain the
properties of the plane wave projected into the spherical sky. The modified
representations are spin $0$ quantities, and they also yield the E-mode and\
B-mode polarization patterns \cite{zaldarriaga1997, kamionkowski1997, Kim}.
They are constructed by means of Eqs. (\ref{amp1}), (\ref{angE1}) and (\ref%
{angB1}), and graphically represented in Figures \ref{fig5}, \ref{fig6} and %
\ref{fig7} (see \textbf{Graphic representations 2}) in Section \ref{plane};
with different conditions of the plane wave projection into the spherical
sky.
 
Overall, GWs can generate the gradient- and curl-like E- and B-modes as expected.
Even though, notable differences between \textbf{Graphic
representations 1 }and\textbf{\ 2} are observed due to the plane wave
modulation; parameterized by $\rho $, which defines the plane wave
projection into the spherical sky. With small to medium $\rho $, which
corresponds to the plane wave modulation at \textit{long} to \textit{medium}
wavelengths or at \textit{small} to \textit{medium} comoving distances to
the last scattering surface, the E-modes and B-modes show broken symmetric
and anti-symmetric patterns, respectively in \textbf{Graphic representations
2}. However, with large $\rho $, which corresponds to the plane wave
modulation at \textit{short} wavelengths or at \textit{large} comoving
distances to the last scattering surface, the E-modes and B-modes take
proper symmetric and anti-symmetric patterns. In fact, in the limit $\rho
\gg 1$, the E-modes and\ B-modes in \textbf{Graphic representations 2} scale
to their counterparts in \textbf{Graphic representations 1}, apart from the
factor $\rho ^{2}\sin ^{2}\theta $ in the amplitude \cite{Chiang}. In both 
\textbf{Graphic representations 1 }and\textbf{\ 2}, however, the periodicity
of the patterns in $\phi $ for the E-modes and B-modes remains the same,
i.e. $\pi /2$.

In this work we have restricted our attention to the case of a single
charged particle under the influence of GWs passing through the CMB. We have
investigated how the EM radiation is rescattered by the charge being
agitated by the GWs and how the rescattered radiation becomes polarized in
the CMB. More realistic astronomical scenarios, however, would involve
multiple charged particles interacting with each other. At the same time, as
we assume that our universe was expanding during the epoch of reionization,
the interaction between the particles would be significantly affected by the
scale factor of the expanding universe. Therefore, to handle a situation
with the interaction properly, the motion of the particles should be
described in the spacetime geometry which is prescribed by the perturbed
Friedmann-Robertson-Walker metric rather than by the perturbed flat
spacetime metric. Unlike the case of a single particle, dynamic evolution of
the multiple particles in this setup should be treated statistically, e.g.
using the Boltzmann equation \cite{zaldarriaga1997, kamionkowski1997, Hu2}.
Likewise, astrophysically realistic CMB polarization maps -- which we do not
provide here -- result from convolution of the patterns caused by
single-particle interaction with the density distribution of the CMB plasma.
Furthermore, we do not yet present power spectra in this study. Following \cite {Hu2}, we can determine the $\ell$-mode expressions for the power spectra out of Eqs. (\ref{QpmU3}) and (\ref{Qt}) -- (\ref{Bt}) by means of the Clebsch-Gordan relation. With these we will be able to estimate the E-mode/B-mode contributions in a quantitative manner as shown by Fig. 11 (b) in \cite{Hu}.
We plan to address these open issues in a follow-up study.

\section*{Acknowledgments}

We are grateful to the anonymous referee for valuable suggestions and comments.
D.-H. Kim acknowledges financial support from the National Research
Foundation of Korea (NRF) via Basic Research Grant NRF-2013R1A1A2008901. S.
Trippe acknowledges financial support from the National Research
Foundation of Korea (NRF) via Basic Research Grant NRF-2015R1D1A1A01056807.
Correspondence should be addressed to S.~T.

\appendix

%\newcounter{equation} \setcounter{equation}{0} \renewcommand{%
%\theequation}{A-\arabic{equation}}

\section{Non-polarized anisotropic incident radiation \label{incident}}

In Subsection \ref{EM} we prescribe an anisotropic incident monochromatic
radiation field on a charge via the two waves,%
\begin{eqnarray}
\mathbf{E}_{\mathrm{I}\,(\mathrm{in})} &=&E_{\mathrm{I}}\left\{ \mathbf{e}%
_{y}\exp \left[ \mathrm{i}\left( Kx-\Omega t+\varphi _{\mathrm{I}}\left(
t\right) \right) +\mathbf{e}_{z}\exp \left[ \mathrm{i}\left( Kx-\Omega
t+\vartheta _{\mathrm{I}}\left( t\right) \right) \right] \right] \right\} ,
\label{a1} \\
\mathbf{E}_{\mathrm{II}\,(\mathrm{in})} &=&E_{\mathrm{II}}\left\{ \mathbf{e}%
_{x}\exp \left[ \mathrm{i}\left( Ky-\Omega t+\varphi _{\mathrm{II}}\left(
t\right) \right) \right] +\mathbf{e}_{z}\exp \left[ \mathrm{i}\left(
Ky-\Omega t+\vartheta _{\mathrm{II}}\left( t\right) \right) \right] \right\}
,  \label{a2}
\end{eqnarray}%
where $E_{\mathrm{I}}$ and $E_{\mathrm{II}}$ ($E_{\mathrm{I}}\neq E_{\mathrm{%
II}}$) are the amplitudes, and $\Omega $ and $K$ are the angular temporal
frequency and angular spatial frequency with $\Omega =cK$, and $\varphi _{%
\mathrm{I}}\left( t\right) $, $\vartheta _{\mathrm{I}}\left( t\right) $, $%
\varphi _{\mathrm{II}}\left( t\right) $ and $\vartheta _{\mathrm{II}}\left(
t\right) $ are the phase modulation functions, which are assumed to vary on
a time scale much slower than the period of the waves, i.e. $\left\vert \dot{%
\varphi}_{\mathrm{I}}\right\vert $, $\left\vert \dot{\vartheta}_{\mathrm{I}%
}\right\vert $, $\left\vert \dot{\varphi}_{\mathrm{II}}\right\vert $, $%
\left\vert \dot{\vartheta}_{\mathrm{II}}\right\vert \ll \Omega $.

With this prescription, $\mathbf{E}_{\mathrm{I}\,(\mathrm{in})}$ and $%
\mathbf{E}_{\mathrm{II}\,(\mathrm{in})}$ are \textit{unpolarized}. This is
due to the relative phases $\varphi _{\mathrm{I}}\left( t\right) -\vartheta
_{\mathrm{I}}\left( t\right) $ and $\varphi _{\mathrm{II}}\left( t\right)
-\vartheta _{\mathrm{II}}\left( t\right) $ which fluctuate in time and will
not let each wave remain in a single polarization state. This can be shown
formally by computing the Stokes parameters out of the waves (\ref{a1}) and (%
\ref{a2}). By Eqs. (\ref{I0}) -- (\ref{V0}) we calculate 
\begin{eqnarray}
I_{(\mathrm{I})} &=&\left\langle \left\vert E_{\mathrm{I~}y\,(\mathrm{in}%
)}\right\vert ^{2}\right\rangle +\left\langle \left\vert E_{\mathrm{I~}z\,(%
\mathrm{in})}\right\vert ^{2}\right\rangle ,  \label{a3} \\
Q_{(\mathrm{I})} &=&\left\langle \left\vert E_{\mathrm{I~}y\,(\mathrm{in}%
)}\right\vert ^{2}\right\rangle -\left\langle \left\vert E_{\mathrm{I~}z\,(%
\mathrm{in})}\right\vert ^{2}\right\rangle ,  \label{a4} \\
U_{(\mathrm{I})} &=&\left\langle E_{\mathrm{I~}y\,(\mathrm{in})}E_{\mathrm{I~%
}z\,(\mathrm{in})}^{\ast }\right\rangle +\left\langle E_{\mathrm{I~}y\,(%
\mathrm{in})}^{\ast }E_{\mathrm{I~}z\,(\mathrm{in})}\right\rangle ,
\label{a5} \\
V_{(\mathrm{I})} &=&\mathrm{i}\left( \left\langle E_{\mathrm{I~}y\,(\mathrm{%
in})}E_{\mathrm{I~}z\,(\mathrm{in})}^{\ast }\right\rangle -\left\langle E_{%
\mathrm{I~}y\,(\mathrm{in})}^{\ast }E_{\mathrm{I~}z\,(\mathrm{in}%
)}\right\rangle \right) ,  \label{a6}
\end{eqnarray}%
and%
\begin{eqnarray}
I_{(\mathrm{II})} &=&\left\langle \left\vert E_{\mathrm{II~}x\,(\mathrm{in}%
)}\right\vert ^{2}\right\rangle +\left\langle \left\vert E_{\mathrm{II~}z\,(%
\mathrm{in})}\right\vert ^{2}\right\rangle ,  \label{a7} \\
Q_{(\mathrm{II})} &=&\left\langle \left\vert E_{\mathrm{II~}x\,(\mathrm{in}%
)}\right\vert ^{2}\right\rangle -\left\langle \left\vert E_{\mathrm{II~}z\,(%
\mathrm{in})}\right\vert ^{2}\right\rangle ,  \label{a8} \\
U_{(\mathrm{II})} &=&\left\langle E_{\mathrm{II~}x\,(\mathrm{in})}E_{\mathrm{%
II~}z\,(\mathrm{in})}^{\ast }\right\rangle +\left\langle E_{\mathrm{II~}x\,(%
\mathrm{in})}^{\ast }E_{\mathrm{II~}z\,(\mathrm{in})}\right\rangle ,
\label{a9} \\
V_{(\mathrm{II})} &=&\mathrm{i}\left( \left\langle E_{\mathrm{II~}x\,(%
\mathrm{in})}E_{\mathrm{II~}z\,(\mathrm{in})}^{\ast }\right\rangle
-\left\langle E_{\mathrm{II~}x\,(\mathrm{in})}^{\ast }E_{\mathrm{II~}z\,(%
\mathrm{in})}\right\rangle \right) .  \label{a10}
\end{eqnarray}%
From Eqs. (\ref{a1}) and (\ref{a2}) we easily find $\left\langle \left\vert
E_{\mathrm{I~}y\,(\mathrm{in})}\right\vert ^{2}\right\rangle =\left\langle
\left\vert E_{\mathrm{I~}z\,(\mathrm{in})}\right\vert ^{2}\right\rangle =E_{%
\mathrm{I}}^{2}$ and $\left\langle \left\vert E_{\mathrm{II~}x\,(\mathrm{in}%
)}\right\vert ^{2}\right\rangle =\left\langle \left\vert E_{\mathrm{II~}z\,(%
\mathrm{in})}\right\vert ^{2}\right\rangle =E_{\mathrm{II}}^{2}$. However,
for the cross-terms we find 
\begin{eqnarray}
\left\langle E_{\mathrm{I~}y\,(\mathrm{in})}E_{\mathrm{I~}z\,(\mathrm{in}%
)}^{\ast }\right\rangle &=&\frac{E_{\mathrm{I}}^{2}}{T}\int_{0}^{T}dt~e^{%
\mathrm{i}\left[ \varphi _{\mathrm{I}}\left( t\right) -\vartheta _{\mathrm{I}%
}\left( t\right) \right] },  \label{a11} \\
\left\langle E_{\mathrm{I~}y\,(\mathrm{in})}^{\ast }E_{\mathrm{I~}z\,(%
\mathrm{in})}\right\rangle &=&\frac{E_{\mathrm{I}}^{2}}{T}\int_{0}^{T}dt~e^{-%
\mathrm{i}\left[ \varphi _{\mathrm{I}}\left( t\right) -\vartheta _{\mathrm{I}%
}\left( t\right) \right] },  \label{a12}
\end{eqnarray}%
and%
\begin{eqnarray}
\left\langle E_{\mathrm{II~}x\,(\mathrm{in})}E_{\mathrm{II~}z\,(\mathrm{in}%
)}^{\ast }\right\rangle &=&\frac{E_{\mathrm{II}}^{2}}{T}\int_{0}^{T}dt~e^{%
\mathrm{i}\left[ \varphi _{\mathrm{II}}\left( t\right) -\vartheta _{\mathrm{%
II}}\left( t\right) \right] },  \label{a13} \\
\left\langle E_{\mathrm{II~}x\,(\mathrm{in})}^{\ast }E_{\mathrm{II~}z\,(%
\mathrm{in})}\right\rangle &=&\frac{E_{\mathrm{II}}^{2}}{T}%
\int_{0}^{T}dt~e^{-\mathrm{i}\left[ \varphi _{\mathrm{II}}\left( t\right)
-\vartheta _{\mathrm{II}}\left( t\right) \right] }.  \label{a14}
\end{eqnarray}%
For a sufficiently long time $T$, the phases $\pm \left[ \varphi _{\mathrm{I}%
}\left( t\right) -\vartheta _{\mathrm{I}}\left( t\right) \right] $ and $\pm %
\left[ \varphi _{\mathrm{II}}\left( t\right) -\vartheta _{\mathrm{II}}\left(
t\right) \right] $ will fluctuate randomly and thus their trigonometric
evaluations will be uniformly distributed over $\left[ -1,1\right] $ for $%
t\in \left[ 0,T\right] $. This will lead the integrals $\int_{0}^{T}dt~e^{%
\pm \mathrm{i}\left[ \varphi _{\mathrm{I}}\left( t\right) -\vartheta _{%
\mathrm{I}}\left( t\right) \right] }$ and $\int_{0}^{T}dt~e^{\pm \mathrm{i}%
\left[ \varphi _{\mathrm{II}}\left( t\right) -\vartheta _{\mathrm{II}}\left(
t\right) \right] }$ to be finite and hence $\int_{0}^{T}dt~e^{\pm \mathrm{i}%
\left[ \varphi _{\mathrm{I}}\left( t\right) -\vartheta _{\mathrm{I}}\left(
t\right) \right] }/T$ and $\int_{0}^{T}dt~e^{\pm \mathrm{i}\left[ \varphi _{%
\mathrm{II}}\left( t\right) -\vartheta _{\mathrm{II}}\left( t\right) \right]
}/T$ to be vanishingly small. Therefore, we finally have%
\begin{equation}
\left[ 
\begin{array}{c}
I_{(\mathrm{I,II})} \\ 
Q_{(\mathrm{I,II})} \\ 
U_{(\mathrm{I,II})} \\ 
V_{(\mathrm{I,II})}%
\end{array}%
\right] =2E_{\mathrm{I,II}}^{2}\left[ 
\begin{array}{c}
1 \\ 
0 \\ 
0 \\ 
0%
\end{array}%
\right] ,  \label{a15}
\end{equation}%
which was to be proved.

%\newcounter{equation} \setcounter{equation}{0} \renewcommand{%
%\theequation}{B-\arabic{equation}}

\section{Solving the geodesic equation of motion for a charge in perturbed
spacetime \label{solve}}

In Section \ref{alongz} the geodesic equation of motion for a charge in the
spacetime $g_{\mu \nu }=\eta _{\mu \nu }+h_{\mu \nu }$ in linearized gravity
reads%
\begin{equation}
\ddot{X}^{i}+\eta ^{ik}h_{jk,t}\dot{X}^{j}+\frac{1}{2}\eta ^{il}\left(
h_{jl,k}+h_{kl,j}-h_{jk,l}\right) \dot{X}^{j}\dot{X}^{k}=\frac{q}{m}\left(
\eta ^{ik}-h_{~}^{ik}\right) E_{k},  \label{b1}
\end{equation}%
where the GWs $h_{ij}$ are given by 
\begin{eqnarray}
h_{ij}^{+} &=&h\left( e_{i}^{1}\otimes e_{j}^{1}-e_{i}^{2}\otimes
e_{j}^{2}\right) \exp \left[ \mathrm{i}\omega \left( \frac{z}{c}-t\right) %
\right] ,  \label{b2} \\
h_{ij}^{\times } &=&h\left( e_{i}^{1}\otimes e_{j}^{2}+e_{i}^{2}\otimes
e_{j}^{1}\right) \exp \left[ \mathrm{i}\omega \left( \frac{z}{c}-t\right) %
\right] ,  \label{b3}
\end{eqnarray}%
with $h$ representing the strain amplitude, $\omega $ the frequency and $+$/$%
\times $ the two polarization states prescribed by the tensors $%
e_{i}^{1}\otimes e_{j}^{1}-e_{i}^{2}\otimes e_{j}^{2}$ and $e_{i}^{1}\otimes
e_{j}^{2}+e_{i}^{2}\otimes e_{j}^{1}$, and on the right-hand side the EM
field is given by%
\begin{equation}
E_{i}=e_{i}^{2}E_{\mathrm{I}}\exp \left[ \mathrm{i}\left( \Omega \left( 
\frac{z}{c}-t\right) +\varphi _{\mathrm{I}}\left( t\right) \right) \right]
+e_{i}^{1}E_{\mathrm{II}}\exp \left[ \mathrm{i}\left( \Omega \left( \frac{z}{%
c}-t\right) +\varphi _{\mathrm{II}}\left( t\right) \right) \right] ,
\label{b4}
\end{equation}%
which describes the outgoing field reflected by the charge.

The geodesic equation (\ref{b1}) can be solved via perturbation. First, the
unperturbed part of a solution for $\dot{X}^{i}$=$\left( \dot{x},\dot{y},%
\dot{z}\right) $, namely $\dot{X}_{\left[ 0\right] }^{i}=\left( \dot{x}_{%
\left[ 0\right] },\dot{y}_{\left[ 0\right] },\dot{z}_{\left[ 0\right]
}\right) $ is obtained as follows. Plugging $\dot{X}^{i}=\dot{X}_{\left[ 0%
\right] }^{i}$ into Eq. (\ref{b1}), and keeping only the unperturbed terms,
we have 
\begin{equation}
\ddot{X}_{\left[ 0\right] }^{i}=\frac{q}{m}\eta ^{ik}E_{k}.  \label{b5}
\end{equation}%
Inserting Eq. (\ref{b4}) into Eq. (\ref{b5}), and integrating this over $t$, 
\begin{eqnarray}
\dot{X}_{\left[ 0\right] }^{i} &=&v_{\mathrm{o}}^{i}+\frac{q}{m}\eta
^{ik}\int_{0}^{t}dt^{\prime }E_{k}  \notag \\
&=&v_{\mathrm{o}}^{i}+\frac{qE_{\mathrm{o}i}}{m}\int_{0}^{t}dt^{\prime }\exp %
\left[ \mathrm{i}\left( \Omega \left( \frac{z}{c}-t^{\prime }\right)
+\varphi _{i}\left( t^{\prime }\right) \right) \right] ,  \label{b6}
\end{eqnarray}%
where $E_{\mathrm{o}i}=\left( E_{\mathrm{II}},E_{\mathrm{I}},0\right) $ and $%
\varphi _{i}\left( t\right) =\left( \varphi _{\mathrm{II}}\left( t\right)
,\varphi _{\mathrm{I}}\left( t\right) ,0\right) $. Now, we take the
time-dependent part from the integral in Eq. (\ref{b6}) and define 
\begin{equation}
\mathcal{I}\left( t\right) \equiv \int_{0}^{t}dt^{\prime }e^{-\mathrm{i}%
\left[ \Omega t^{\prime }-\varphi \left( t^{\prime }\right) \right] },
\label{b7}
\end{equation}%
where $\varphi \left( t\right) $ refers to either $\varphi _{\mathrm{I}%
}\left( t\right) $ or $\varphi _{\mathrm{II}}\left( t\right) $. Using
integration by parts repeatedly, we arrive at%
\begin{eqnarray}
\mathcal{I}\left( t\right) &=&\mathrm{i}\frac{e^{-\mathrm{i}\left[ \Omega
t-\varphi \left( t\right) \right] }}{\Omega }\left\{ 1+\frac{\dot{\varphi}%
\left( t\right) }{\Omega }+\left[ \left( \frac{\dot{\varphi}\left( t\right) 
}{\Omega }\right) ^{2}-\mathrm{i}\frac{\ddot{\varphi}\left( t\right) }{%
\Omega ^{2}}\right] +\left[ \left( \frac{\dot{\varphi}\left( t\right) }{%
\Omega }\right) ^{3}-3\mathrm{i}\frac{\dot{\varphi}\left( t\right) }{\Omega }%
\frac{\ddot{\varphi}\left( t\right) }{\Omega ^{2}}-\frac{\dddot{\varphi}%
\left( t\right) }{\Omega ^{3}}\right] +\cdots \right\}  \notag \\
&=&\mathrm{i}\frac{e^{-\mathrm{i}\left[ \Omega t-\varphi \left( t\right) %
\right] }}{\Omega }+\mathcal{O}_{\left[ 0\right] }\left( \dot{\varphi}%
/\Omega ,\ddot{\varphi}/\Omega ^{2},\dot{\varphi}\ddot{\varphi}/\Omega ^{3},%
\dddot{\varphi}/\Omega ^{3},\cdots \right)  \label{b8}
\end{eqnarray}%
where $\dot{\varphi}/\Omega \sim \varepsilon $, $\ddot{\varphi}/\Omega
^{2}\sim \varepsilon ^{2}$, $\dot{\varphi}\ddot{\varphi}/\Omega ^{3}\sim 
\dddot{\varphi}/\Omega ^{3}\sim \varepsilon ^{3}$, etc. with $\varepsilon
\ll 1$, from the assumption that $\varphi \left( t\right) $ varies on a time
scale much slower than the period of the wave, i.e. $\left\vert \dot{\varphi}%
\right\vert $ $\ll \Omega $, $\left\vert \ddot{\varphi}\right\vert \ll
\Omega ^{2}$, $\left\vert \dddot{\varphi}\right\vert \ll \Omega ^{3}$, etc.
This leads to%
\begin{equation}
\dot{X}_{\left[ 0\right] }^{i}=v_{\mathrm{o}}^{i}+\mathrm{i}\frac{qE_{%
\mathrm{o}i}}{m\Omega }\exp \left[ \mathrm{i}\left( \Omega \left( \frac{z}{c}%
-t\right) +\varphi _{i}\left( t\right) \right) \right] +\mathcal{O}_{\left[ 0%
\right] }\left( \dot{\varphi}_{i}/\Omega ,\ddot{\varphi}_{i}/\Omega ^{2},%
\dot{\varphi}_{i}\ddot{\varphi}_{i}/\Omega ^{3},\dddot{\varphi}_{i}/\Omega
^{3},\cdots \right) .  \label{b9}
\end{equation}

A first-order perturbation of $\dot{X}^{i}$, namely $\dot{X}_{\left[ h\right]
}^{i}$ is obtained by recycling the unperturbed part, i.e. Eq. (\ref{b9})
into Eq. (\ref{b1}). Plugging $\dot{X}^{i}=\dot{X}_{\left[ 0\right] }^{i}+%
\dot{X}_{\left[ h\right] }^{i}$ into Eq. (\ref{b1}), and keeping only
first-order terms in $h$,%
\begin{equation}
\ddot{X}_{\left[ h\right] }^{i}=-\frac{q}{m}h_{~}^{ik}E_{k}-\eta
^{ik}h_{jk,t}\dot{X}_{\left[ 0\right] }^{j}-\frac{1}{2}\eta ^{il}\left(
h_{jl,k}+h_{kl,j}-h_{jk,l}\right) \dot{X}_{\left[ 0\right] }^{j}\dot{X}_{%
\left[ 0\right] }^{k}.  \label{b10}
\end{equation}%
Integrating this over $t$,%
\begin{equation}
\dot{X}_{[h]}^{i}=-\frac{q}{m}\int_{0}^{t}dt^{\prime }h_{~}^{ik}E_{k}-\eta
^{ik}\int_{0}^{t}dt^{\prime }h_{jk,t^{\prime }}\dot{X}_{\left[ 0\right]
}^{j}-\frac{1}{2}\eta ^{il}\int_{0}^{t}dt^{\prime }\left(
h_{jl,k}+h_{kl,j}-h_{jk,l}\right) \dot{X}_{\left[ 0\right] }^{j}\dot{X}_{%
\left[ 0\right] }^{k},  \label{b11}
\end{equation}%
where the integrands are specified by Eqs. (\ref{b2}), (\ref{b3}), (\ref{b4}%
) and (\ref{b9}). Here each integral has the same form as Eq. (\ref{b7}) and
hence can be approximated in the same manner as in Eq. (\ref{b8}), thereby
generating the error $\mathcal{O}_{\left[ h\right] }\left( \dot{\varphi}%
/\Omega ,\ddot{\varphi}/\Omega ^{2},\dot{\varphi}\ddot{\varphi}/\Omega ^{3},%
\dddot{\varphi}/\Omega ^{3},\cdots \right) \sim h\times \mathcal{O}_{\left[ 0%
\right] }\left( \dot{\varphi}/\Omega ,\ddot{\varphi}/\Omega ^{2},\dot{\varphi%
}\ddot{\varphi}/\Omega ^{3},\dddot{\varphi}/\Omega ^{3},\cdots \right) $.
Combining the result from Eq. (\ref{b11}) with Eq. (\ref{b9}), we obtain the
full first-order perturbation solution, i.e. $\dot{X}^{i}=\dot{X}_{\left[ 0%
\right] }^{i}+\dot{X}_{\left[ h\right] }^{i}$, which has the total error $%
\mathcal{O}_{\left[ 0\right] }\left( \dot{\varphi}/\Omega ,\ddot{\varphi}
/\Omega ^{2},\dot{\varphi}\ddot{\varphi}/\Omega ^{3},\dddot{\varphi}/\Omega
^{3},\cdots \right) +\mathcal{O}_{\left[ h\right] }\left( \dot{\varphi}
/\Omega ,\ddot{\varphi}/\Omega ^{2},\dot{\varphi}\ddot{\varphi}/\Omega ^{3},
\dddot{\varphi}/\Omega ^{3},\cdots \right) $, as presented in Eqs. (\ref{px}
) -- (\ref{cz}).

The full first-order perturbation solution for $\ddot{X}^{i}$, i.e. $\ddot{X}
^{i}=\ddot{X}_{\left[ 0\right] }^{i}+\ddot{X}_{\left[ h\right] }^{i}$ can be
obtained by combining Eqs. (\ref{b5}) and (\ref{b10}). The unperturbed part $
\ddot{X}_{\left[ 0\right] }^{i}$ is trivially obtained from the right-hand
side of (\ref{b5}), which is equivalent to $E_{i}$ (\ref{b4}) apart from the
factor; thereby not generating $\mathcal{O}_{\left[ 0\right] }$. However,
the perturbed part $\ddot{X}_{\left[ h\right] }^{i}$ is obtained by
recycling the unperturbed part $\dot{X}_{\left[ 0\right] }^{i}$ (\ref{b9})
into (\ref{b10}); thereby generating the error $\mathcal{O}_{\left[ h\right]
}\sim h\times \mathcal{O}_{\left[ 0\right] }$. Hence the total error from $%
\ddot{X}^{i}=\ddot{X}_{\left[ 0\right] }^{i}+\ddot{X}_{\left[ h\right] }^{i}$
is $\mathcal{O}_{\left[ h\right] }$ only. The rescattered EM radiation is
obtained by 
\begin{equation}
E_{\,(\mathrm{scat})}^{i}=\frac{q\ddot{X}^{i}}{r}=\frac{q}{r}\left( \ddot{X}%
_{\left[ 0\right] }^{i}+\ddot{X}_{\left[ h\right] }^{i}\right) ,  \label{b12}
\end{equation}%
due to Eqs. (\ref{Ex}), (\ref{Ey}) and (\ref{Ez}), and therefore it has the
error $\mathcal{O}_{\left[ h\right] }$ only, as shown by Eqs. (\ref{Epx}) - (%
\ref{Ecz}).

%%% REFERENCES %%%%%%%%%%%%%%%%%%%%%%%%%%%%%%%%%%%%%%%%%%%%%%%%%%%%%%

\end{document}